
\input harvmac
\def\u{{\bf u}}
\def\U{{\bf U}}
\def\V{{\bf V}}
\def\S{{\cal S}}
\def\H{{\bf H}}

\def\M{{\cal M}}
\def\vv{{\bf v}}
\def\h{{\bf h}}
\def\CS{{\cal S}}

\def\CN{{\cal N}}

\font\cmss=cmss10 \font\cmsss=cmss10 at 7pt
\def\IZ{\relax\ifmmode\lrefhchoice
{\hbox{\cmss Z\kern-.4em Z}}{\hbox{\cmss Z\kern-.4em Z}}
{\lower.9pt\hbox{\cmsss Z\kern-.4em Z}}
{\lower1.2pt\hbox{\cmsss Z\kern-.4em Z}}\else{\cmss Z\kern-.4em Z}}
\def\np{Nucl. Phys. }
\def\pl{Phys. Lett. }
\def\pr{Phys. Rev. }
\def\prl{Phys. Rev. Lett. }

\def\CF{{\cal F}}
\def \CO{{\cal O}}

\def \CA{{\cal A}}

\def \CM{{\cal M}}
\def \CA{{\cal A}}
\def \CE{{\cal E}}

\def \l{\ell}

\def \sinh{{\rm sinh}}
\def \cosh{{\rm cosh}}
\def\I{{\bf I}}
\def\CD {{\cal D}}

\def\CL {{\cal L}}

\def\CO {{\cal O}}
\def\p {\partial}
\def\CS {{\cal S}}
\font\cmss=cmss10 \font\cmsss=cmss10 at 7pt
\def\IZ{\relax\ifmmode\mathchoice
{\hbox{\cmss Z\kern-.4em Z}}{\hbox{\cmss Z\kern-.4em Z}}
{\lower.9pt\hbox{\cmsss Z\kern-.4em Z}}
{\lower1.2pt\hbox{\cmsss Z\kern-.4em Z}}\else{\cmss Z\kern-.4em Z}}

\def\l {\ell }

\def\a{\alpha}
\def\ta{\tilde \alpha}

\def\tb{\tilde \beta}

\def\tv{\tilde \beta}
\def\v{\beta}
\def\tr{ {\rm tr}}
\def\p {\partial}
\def\CS {{\cal S}}
\def\CL {{\cal L}}
\def\CF {{\cal F}}
\def\CZ{{\cal Z}}
\def\CA {{\cal A}}

\def\CD {{\cal D}}
\def\pl{{\it Phys. Lett. } }
\def\np{{\it Nucl. Phys.} }
\def\pr{{\it Phys. Rev.} }
\def\prl{{\it Phys. Rev. Lett.} }

\def\R{\relax{\rm I\kern-.18em R}}
\font\cmss=cmss10 \font\cmsss=cmss10 at 7pt
\def\Z{\relax\ifmmode\mathchoice
{\hbox{\cmss Z\kern-.4em Z}}{\hbox{\cmss Z\kern-.4em Z}}
{\lower.9pt\hbox{\cmsss Z\kern-.4em Z}}
{\lower1.2pt\hbox{\cmsss Z\kern-.4em Z}}\else{\cmss Z\kern-.4em Z}}
\def\C{{\cal C}}


\lref\notrn{A. M. Polyakov, \pl 72B (1978) 447;
J.-M. Drouffe, \pr D18 (1978) 1174;
L. Susskind, \pr D20 (1979) 2610;
M. Nauenberg and D. Toussaint, \np B190 (1981) 288;
P. Menotti and E. Onofri, \np B190 (1984) 288;
C.B. Lang, P. Salomonson and B.S. Skagerstam, \pl 107B (1981) 211,
\np B190 (1981) 337}.
 \lref\ros{P. Rossi, {\it Ann. Phys.}
132 (1981) 463}
\lref\mm{Yu. M. Makeenko and A. A. Migdal, \np B188 (1981) 269,
A. A. Migdal, \pr 102 (1983) 201}
\lref\tho{G. 't Hooft, \np B72 (1974) 461}.

\lref\rus{B. Rusakov, {\it Mod. Phys. Lett.} A5 (1990) 693}
\lref\gt{ D. Gross, \np B400 (1993) 161;
 D. Gross and W. Taylor IV, \np B400 (1993) 181}
 \lref\dg{ M. Douglas, Preprint RU-93-13 (NSF-ITP-93-39);
J. Minahan and A. Polychronakos, hep-th/9303153}
\lref\dgk{ M. Douglas and V. Kazakov, preprint LPTENS-93/20}

\Title{}
{\vbox{\centerline
{$U(N)$ Gauge Theory and Lattice Strings   }
\vskip2pt
\centerline{
 }}}

\vskip4pt

\centerline{Ivan K. Kostov \footnote{$ ^\ast $}{on leave of absence
from the Institute for Nuclear Research and Nuclear Energy,
Boulevard Tsarigradsko Chauss\'ee 72, BG-1784 Sofia, Bulgaria}
\footnote{$^{\# }$}{(kostov@amoco.saclay.cea.fr)}
}
\centerline{{\it Service de Physique Th\'eorique
\footnote{$ ^\dagger$}{Laboratoire de la Direction des Sciences
de la Mati\`ere du Comissariat \`a l'Energie Atomique} de Saclay
CE-Saclay, F-91191 Gif-Sur-Yvette, France}}

\vskip .3in

\baselineskip12pt{
The    $U(N)$ gauge theory  on a
$D$-dimensional lattice is reformulated
as a theory of lattice strings (a statistical model of  random surfaces).
The Boltzmann weights of the surfaces can have both signs and are tuned
so that the longitudinal modes of the string are elliminated.
The $U(\infty)$ gauge theory is described by noninteracting planar
surfaces and the  $1/N$ corrections are produced by
surfaces with higher topology as well as by contact interactions due to
microscopic tubes, trousers, handles, etc.
We pay special attention to the case $D=2$ where the sum over surfaces
can be performed explicitly, and demonstrate that it reproduces the known
exact results for the free energy and  Wilson loops in the continuum limit.
In $D=4$ dimensions, our lattice string model reproduces the
strong coupling phase of the gauge theory.
The weak coupling phase
 is described by a more complicated string
whose  world surface may have windows.
A possible integration
measure in the space of continuous surfaces is suggested.

}

\vskip 1cm
\leftline{{ \it Submitted to Nuclear Physics B}}
\smallskip
\Date{ September  1993 }

\baselineskip=14pt plus 2pt minus 2pt

\newsec{Introduction}

In the standard perturbative formulation of QCD the elementary
excitations are coloured pointlike particles - gluons.
At very small distances   the gluons are asymptotically free
while at distances comparable with the correlation length
of the theory (the typical hadron radius), the effective interaction
becomes so strong that the very concept of gluons looses its
sense.
The relevant degrees of freedom for the infrared regime of QCD
are expected to be
colourless extended objects (strings).
The corresponding operators
  are the loop fields
\eqn\plaq{\Phi (\C)=
{\tr \over N}\ {\rm P}\  \exp[i\oint _{\C}dx_{\mu}A_{\mu}(x)]
 }
are defined in the space of all closed contours $\C$.
The letter P in front of the exponential stands for Dyson ordering along
the contour.

 As it was pointed out by 't Hooft
\tho , the perturbative expansion in the $N$-colour quantum chromodynamics
has the same topological structure as the loop expansion in a string field
theory with coupling constant $1/N$\foot{Note that, unlike the usual string
theories, the coupling constant is a dimensionless number and
does not renormalize.}.
One can  imagine the world surface of the  chromodynamical string as
the result of condensation of dense planar diagrams.
Then in the large $N$ limit
the Wilson loop average $W(\C)= \langle \Phi (\C) \rangle $
would be given by the
sum over all planar
 surfaces $\S$ spanning the loop $\C$, with some weight $\Omega (\S)$.
Symbolically
\eqn\iij{W (\C)= \sum _{\p \S = \C} \Omega (\S).}

The perturbation theory does not tell  much about the form of the
Boltzmann factor $\Omega (\S)$ but it is known  \mm\ that the
 functional $W(\C)$ satisfies a  loop equation involving a local deformation
of the contour and having a local
(vector)  contact term.
Therefore it is very plausible that there exists a
 string representation
of the Wilson loop average of the type \iij\ in which  the
 weight of each  surface depends
on the geometry of the world sheet through
a local action.

The loop functional \iij\ can be determined in principle from
the following conditions.
First, it does not change if a backtracking piece $\Gamma \Gamma ^{-1}$
is added to the contour $\C$
\eqn\unn{W(\C \Gamma \Gamma ^{-1})=W(\C)}
For such functionals ( called in \mm\
Stokes type functionals),  one can define the
 so-called area derivative
$\delta /\delta \sigma _{\mu \nu}$ defined by
\eqn\ardv{\delta W(\C)= \oint _{\C} (dx_{\mu}\delta x_{\nu}-
dx_{\nu}\delta x_{\mu}) {\delta W(\C) \over
\delta \sigma _{\mu \nu}(x)}}
The equation of motion  for the gauge field
\eqn\eqmt{\nabla _{\mu} F_{\mu \nu}(x)= {\lambda \over N}
 {\delta \over \delta A_{\nu}(x)}, \ \ \  \lambda = N g^2
}
is transformed in the multicolour limit $N\to \infty$
 to the following closed equation for the loop
functional \iij\ derived by Makeenko and Migdal \mm\
\eqn\mme{{\p \over \p x_{\mu}}
 {\delta W(\C) \over
\delta \sigma _{\mu \nu}(x)} = \lambda \oint dy_{\nu} W(\C_{xy})W(\C_{yx})}
It was shown in \mm\ that the iteration of  eq. \mme\
reproduces the series of planar diagrams, in its nonrenormalized form.
 Therefore this equation may help in principle to choose the right
string theory for QCD.

Unfortunately,  eqs. \unn\ and \mme\ are not of much practical use
because they are written for  regularized but not renormalized quantities.
Therefore they imply a definite regularization of the corresponding
string theory which may not be  the simplest one
\foot{In the trivial case $D=2$
there are no divergencies and the renormalized form of the loop
equation \mme\ can be readily found and solved \ref\kk{V.Kazakov and I. Kostov,
\np B176 (1980) 199, V. Kazakov, \np B179 (1981) 283}.
The solution
was  interpreted
as  a sum over minimal surfaces
spanning the loop.}.

The way out of this impass might be found by considering the lattice
definition  of the $U(N)$ gauge theory  \ref\wils{K.
Wilson, \pr D 10 (1974) 2445; A. M. Polyakov, $unpublished$}.
 The lattice formalism, in which the continuous Euclidean spacetime is
replaced by a $D$-dimensional simplicial complex, provides an
unambiguous definition of both gauge theories and random surfaces.
Once the lattice version of the string {\it Ansatz} \iij\ is knowm,
one can try to formulate the world sheet dynamics of the QCD string.

Some years ago, a mechanism of generation of strings within the
strong coupling expansion in $U(N)$ gauge theory
was  suggested by V. Kazakov
\ref\kaza{V.Kazakov, \pl 128B (1983) 316, JETP (Russian
edition) 85 (1983) 1887}.
The idea of  \kaza , borrowed from the Stanley's
 solution \ref\st{H.E. Stanley,
\pr 176 (1968) 718} of the
large-$N$ vector model,
  was to reformulate the $U(N)$ lattice gauge theory
as a kind of  Weingarten theory with additional interactions.
The Weingarten model \ref\wein{D. Weingarten,
\pl 90B (1980)285}
is a simple (but
rather pathological)
gauge theory which is equivalent to a  model of noninteracting
 random surfaces in the large $N$ limit.
In the $U(N)$ lattice gauge theory one can
 impose the unitarity   of the link variable
$\U_{\l}$ by means of a Lagrange multiplier
 and  replace the Haar measure on the
$U(N)$ group by a Gaussian  measure over complex matrices.
 The construction suggested in  \kaza\ has been accomplished in refs.
\ref\obrz{K.H. O'Brien and J.-B. Zuber, \np B253 (1985) 621,
\pl 144B (1984) 407}, \ref\kos{I. Kostov, \pl
138B (1984) 191, 147B (1984) 445}.
In these papers
it was shown that the strong coupling
 expansion of the Wilson $U(\infty)$ lattice gauge
 theory  can be formulated in terms of a model of
 {\it noninteracting} surfaces on the lattice.

Are  these microscopic  strings related to the strings expected to
 the strings expected to
 to arise in  the continuum theory?
Since the weak and the strong coupling phases in $D=4$ dimensions
 are separated by
a first order phase transition, the mechanism of formation of strings
might be very different in the two phases.
We argued
in ref. \ref\ivn{I. Kostov, \np B265 (1986) 223}
  that the strong coupling lattice surfaces and the Feynman-'t Hooft
planar diagrams correspond to two extreme cases of the same generalized model
of random surfaces.
The geometrical meaning of the phase transition
is that
 in the weak coupling phase,  windows (holes with free boundaries)
may appear spontaneously  on the world sheet.
In the continuum limit the world sheet of the string is tore into a
network of thin strips which can be interpreted as gluon propagators.
In this way  the weak coupling phase
is achieved by relaxing the kinematical restrictions obeyed by the
surfaces on the lattice.
Therefore the strong coupling lattice string is perhaps not completely
irrelevant to the continuum limit of the gauge theory.

In this paper we develop further the approach outlined in \ivn .
We derive the string representation of the $U(\infty) $ gauge
theory
using a more traditional mathematical formalism
known from
the mean field analysis in gauge theories
 \ref\jb{J.-M. Drouffe and J.-B. Zuber, {\it Phys. Rep.}
102, Nos. 1,2 (1983)1-119, section 4}.
This formalism
allows to construct
the
 $1/N$ expansion in a systematic way.
The leading term  is given  by the
path integral of the free  string, that is, by the sum over noninteracting
random surfaces on the lattice.
The intrinsic geometry of the world sheet of the   string
allows local singularities (punctures) with curvature $2\pi \times$integer.
These singularities  can occur at  links or cells of the lattice.
The Boltzmann weight of such  surface is equal to
a product of factors associated with the
 punctures and can have positive or negative sign
\foot{This is the reason why the
 triviality theorem of Durhuus, Fr\"ohlich and Jonsson
\ref\dur{B.Durhuus, J. Fr\"ohlich and T. Jonsson, \np B240 (1984) 453}
 is not applicable here.}.
These factors appear as the vacuum expectation values of special
puncture operators located at the links and cells of the lattice.
The world surface is also allowed to make folds but the total contribution of
the surfaces with folds vanishes.

The subleading terms in the $1/N$ expansion have double origin.
First, there is the contribution of the surfaces with
higher topology.
Second, the fluctuations of the puncture operators, which are of
order $1/N^2$,
lead to contact interactions of the  surfaces. The vertices of these
interactions can be imagined as microscopic tubes, trousers, ets. ,
connecting several punctures.

We pay special attention to the case $D=2$ where the equivalence
between the gauge and string theories is established for a
two-dimensional manifold $\M$ with any topology discretized by
a lattice $\CL$.
It happens that in this simplest case this equivalence takes place up to
the continuum limit which is the limit of an infinitely dense lattice
$\CL$.

Very recently, some interesting exact results have been obtained for
the partition function of $U(N)$ and $SU(N)$ gauge theories
defined on a compact two-dimensional surface of genus $G$, refs.
\ref\rus{B. Rusakov, {\it Mod. Phys. Lett.} A5 (1990) 693}
\ref\dg{D. Gross, Princeton preprint PUPT-1356, hep-th/9212149}
\ref\dgwt{D. Gross and W. Taylor IV, preprints PUPT-1376 hep-th/9301068
and PUPT-1382 hep-th/9303046}
\ref\dpdp{M. Douglas, Preprint RU-93-13 (NSF-ITP-93-39);
A. D'Adda, M. Caselle, L. Magnea and S. Panzeri, Preprint hep-th 9304015;
J. Minahan and A. Polychronakos, hep-th/9303153}
\ref\kd{M. Douglas and V. Kazakov, preprint LPTENS-93/20}.
An interpretation of these partition functions in terms of surfaces
 was suggested by Gross and Taylor
\dg , \dgwt . The partition function can be written as a sum
over  oriented coverings of the space-time manifold of   genus $G$ and
area $A$, where different sheets of a covering are glued together by means of
branch points, microscopic
 handles and tubes,  and what they called ``omega points''. The
 ``omega points''
represent topologically  branch points, handles  and tubes but
have no location on the surface. This representation looks particularly
simple on the torus where there are no such
 singular points at all. On a surface
 with genus $G$ there are
$|2-2G| $  ``omega points'' and it is not clear how they can be incorporated
in a local string Lagrangian.

The comparison with the lattice string {\it Ansatz} allows to resolve
 the mystery about the ``omega points'' in the surface
interpretation of the free energy.  Indeed, in the continuum
limit (dense covering lattice $\CL$),   an
 ``omega point'' takes into account
the  contribution of a whole  class of surfaces
distinguished by the position of a {\it localized} singularity (e.g.,
branch point or microscopic tube). The sum of the  Boltzmann weights
of these surfaces is a topological invariant and is not sensitive  to the
choice of the covering lattice $\CL$.
 This is why the ``omega point'' has no fixed location in $\CM$.

The paper is organized as follows.

In sect. 2  we give  a review of
the Stanley's solution \st\
of the
 large-$N$ vector model which may help the reader to
understand the logic of our construction before dipping
into the   technical details.
In sect. 3 we reformulate the functional integral for the gauge theory as an
integral over complex matrix fields with flat measure and regular interaction.
This can be achieved by introducing special puncture operators coupled to
the moments of the matrix fields and  playing
the r\^ole of Lagrange multipliers.  In sect. 4 we perform the
integration over the matrix fields using the 't Hooft diagrammatic rules
and interpret the  result as  the partition
 function of a gas of lattice surfaces
interacting with the vacuum surfaces and among themselves through the puncture
operators. In sect.  5 we study the case of $D=2$ dimensions.
We find
explicitly the weights of the punctures and check the result
on some simple examples of Wilson loops.
Then we consider the
two-loop correlator \kk\  and the free energy on a compact
lattice \kd \  where the contact interactions should be taken into account.
In sect. 6 we discuss the form of the random surface {\it Ansatz}
describing weak coupling phase
and  suggest a measure in the space of continuous surfaces for which
the Wilson loop  functional has well defined area derivative.
In Appendix A  we derive  the microscopic loop equations
for the  puncture operators.

\newsec{The $U(N)$ vector model}
In this section we remind how the  vector field with global  $U(N)$
invariance can be represented in terms of statistical
mechanics of random paths.
 We are using notations maximally close to those to be used later in
the case of the $U(N)$ gauge theory.

The fluctuating field in this model
  is an   $N$-component complex  vector $\u_x=(u_x^1,...,u_x^N)$
defined at the  points $x$ of the $D$-dimensional hypercubical
lattice
and having unit norm
\eqn\nnor{\u^*_x \cdot
\u_x \equiv \sum _{i=1}^{N} |u^i_x|^2 =1}
We assume  a nearest neighbour interaction  $\beta \u^*_x \cdot
\u_y $ associated with
the links $\l = <xy>$ of the lattice.
Then the measure
\eqn\unim{[d\u]= \prod_{i=1}^N d\bar u^i du^i
\ \delta ( \sum _{i=1}^{N} |u^i_x|^2  -1)
= d\u^*d\u \ \delta (\u^*\cdot \u -1) }
 and the interaction are invariant under global $U(N)$
transformations
\eqn\trrt{\u_x \to \U  \u_x; \ \ \U\in U(N)}
The partition function
\eqn\prtf{\CZ =e^{N\CF} = \int \prod _{x}[d\u_x]
\prod _{\l = <xy>} e^{N\beta \u^*_x \cdot \u_{y}}}
can be calculated order by order in the inverse coupling constant $\beta$
by  expanding the exponential as a series of monomials
(strong coupling expansion).
This leads to a diagram technique which in the limit $N\to \infty$
involves tree-like clusters of loops and can be  summed up
explicitly in all orders.

Alternatively one can  transform, by introducing auxiliary fields, the
integral over the original vector field into a duable Gaussian integral
and find an effective
field theory  in which the number $1/N$ plays the r\^ole
of a Planck constant.

 Introducing a new complex vector field
$\h=(h^i,\  i=1,2,...,N)$ we represent the unimodular measure as a flat measure
for the unconstrained complex vector field $\vv=(v^i, i=1,...,N)$.
We insert the exponential representation of the $\delta$-function
\eqn\ddd{\delta (u^j, v^j)= ( {1 \over 2\pi i})^2
\int_{-i\infty}^{i\infty}dh_j d \bar h_j e^{\bar h_j(u^j-v^j)
 +  h_j(\bar u^j- \bar v^j) }}
in the definition \prtf\ of the partition function for every component
$u^j_x$ of the vector field.
Here $\h_{\l}$ and $\vv_{\l}$ are unrestricted complex vectors
on which the integration is performed with a flat measure
\eqn\mmemmm{d\vv = \prod_{j=1}^N dv^j d\bar v^j,\ \ \
d\h = \prod_{j=1}^N dh^j d\bar h^j}
Introfucing the effective potential $F(\h)$ by
\eqn\ffff{e^{NF( \h)}= \int [d\u]e^{N[\u^* \cdot
 \h+\h^*\cdot \u]}}
we write the partition function \prtf\
in the form
\eqn\pppar{
\CZ=e^{N\CF}= \int \prod _{x }\Big(
d\h_{x}d\vv_{x}
e^{-N (\h^*_x \cdot \vv_x + \h_x \cdot \vv_x^*
)}\Big)
\prod_{x} e^{N
F(\h_x)} \prod _{\l = <xx'>}e^{N\beta \vv_x\cdot\vv_x'}}

The potential $F(\h)$ depends only on the radial degree of freedom
\eqn\rrad{\a = \h^{*}\cdot \h}
and can be expanded as
\eqn\sses{
F(\a)=\sum_{n=1}^{\infty}
f_{n}{\alpha^n \over n}
}
 The coefficients of the series are determined by the Ward identity
\eqn\wrdi{(\sum_i N^{-2} \p/\p h^i\p/\p h^*_i - 1)e^{NF}=0}
 or, in
 terms of the $\a$-field,
\eqn\wwwar{({1 \over N}{\p \over \p \a}+\a{1 \over N^2} {\p^2 \over \p \a^2}-1)
e^{NF(\a)}=0}
In the limit $N\to\infty$ the solution is
\eqn\aaaa{
F(\a) =\sqrt{1+4\a}-1-\ln {1+\sqrt{1+4\a} \over 2}
}

Using the integral representation
\eqn\iiii{e^{NF(\h^*\cdot \h)}=\int d\a d\ta e^{N[\ta (\h^*\cdot \h-\a)+
F(\a)]}}
we finally write the partition function
 in a form allowing to perform exactly the integral
over the vector fields
\eqn\npff{\eqalign{
\CZ=e^{N\CF}=&\int \prod_{x}d\a_x d\ta_x\ e^{N[F(\a_x)-\ta_x\a_x]}
 d\vv_x d\h_x
e^{N[-\vv^*_x\cdot \h_x-\h^*_x\cdot \vv_x
]}\cr
&\prod_x  e^{N[\ta _x \h ^*_x\cdot
\h_x ]}
\prod_{\l=<xx'>}e^{N\beta \vv^*_x\cdot \vv_{x'}}\cr}
}
Denoting by $\hat C$
the connectivity matrix of the $D$-dimensional hypercubical   lattice
\eqn\cccac{\hat C=\sum_x \sum_{\mu =1}^{D}
(e^{\p/\p x_\mu}+e^{-\p/\p x_\mu})}
we can write the  result of the   gaussian integration over  the $\vv,\h$
fields
as
\eqn\zzz{e^{N\CF}=\int \prod_{x}d\ta_xd\a_x e^{N[
 F(\a_x ) - \ta_x\a_x]}\ \  e^{N\CF_{0}[\ta]} \equiv
\langle e^{N\CF_0[\ta]}\rangle_{\ta}}
where
\eqn\ffrf{\CF_{0}[\ta]=-\sum_{x} \tr \  \log (1- \beta \ta \hat C)}
is the sum over vacuum loops of the vector fields.

Formally the sum over vacuum loops can be obtained by
expanding the exponential in the last line of \npff\ as a sum of
monomials and applying all possible Wick contractions
\eqn\wwik{\langle \h^{*i}_x \vv_{yj}\rangle = \delta_{x,y} \delta_{j}^{i},
\ \ \ \ \langle \vv^{*i}_x \h_{yj}\rangle = \delta_{x,y} \delta_{j}^{i}}
The vertex $\ta_x \h^*_x\cdot \h_x$  is represented by a node at the point $x$
and contributes a factor $\ta_x$ to the weight of the loop. The
vertex $\beta \vv^*_x\vv_y$ is represented by a segment of the loop
 covering the link
$\l =<xy>$ and contributes a factor $\beta$.
 The weight of each loop is therefore a
 product of local factors  $\beta$ and $\ta_x$
associated with its nodes and links, correspondingly.
Denoting by $\Gamma$ any closed loop on the lattice,
 the r.h.s. of \ffrf\ can be written as
\eqn\spts{\CF_0[h^*]=\sum_{\Gamma}\beta ^{\#{{\rm links}}}
\prod _{x\in \Gamma}\ta_x }

The action for the fields  $\ta , \a $ is proportional to $N$ and in the
limit $N\to \infty$ these fields freeze at their
 expectation values determined by the classical equations of motion
\eqn\eqnm{\ta ={\p F(\a)\over \p \a}={\sqrt{1+4\a}-1 \over 2\a}}
\eqn\eqnmm{\a={\p \CF_0(\ta)\over \p \ta}=
\langle x | {\beta \hat C
\over 1- \ta \beta \hat C} |x\rangle }
or, equivalently,
\eqn\ssas{\ta +\a \ta ^2 =1, \ \ \langle x|
{\ta  \over 1-\beta \ta \hat C}|x\rangle =1}
The last equation has a simple geometrical interpretation in terms of
a random walk on the lattice.
 It can be obtained also directly from the
unimodularity condition $\u^*_x\cdot \u_x=1$.
Consider the two-point Greens function $G_{xy}=\langle \u^*_x\cdot
\u_y\rangle$.
Repeating the same steps as for the partition function we find
\eqn\mnmbm{G_{xy}=\langle \vv^*_x\cdot \vv_y\rangle = \langle x|{\ta \over
1-\beta \hat C}|y \rangle=\sum _{\Gamma_{xy}}\ta ^{\# {\rm nodes}}
\beta^{\#{\rm links}}}
where the sum goes over all paths $\Gamma_{xy}$ between the points $x$ and $y$.
Therefore the second of the equations \ssas\ is just the condition
\eqn\uuuu{G_{xx}=1.}

 After diagonalysing
 the
connectivity matrix we write it as an integral in the momentum space
\eqn\tttr{\int_{-\pi}^{\pi} {1 \over m^2+\sum_{\mu =1}^D[2\sin (p_{\mu}/2)]^2}
=\beta}
\eqn\mmmas{m^2= {1 \over \ta \beta} -2D}
The continuum limit (infinite correlation length) is achieved when
 $m^2\to 0$, i.e.,  $ 2D\beta \ta =1$.

When
 $D\le 2$, the
integral is infrared divergent when $m\to 0$ and gives
\eqn\divv{ \beta =\int {d^Dp \over (2\pi)^D} {1 \over p^2+m^2}
= m^{D-2}, \ \ D<2}

In the marginal case $D=2$ we find the asymptotic freedom law for the
dependence
of the
renormalized coupling $\beta = \beta (m)$ as a function of the
correlation length $1/m$
\eqn\assf{\beta = {1 \over4\pi} \ln {1 \over m^2}}

Above two dimensions the integral is convergent at $m\to 0$ and
the limit of zero mass is achieved at finite value of $\beta = \beta _c$.
  In the weak coupling phase $\beta >\beta _c$ the equations of
motion  \tttr\ have no solution which means that something in our
construction is wrong.
This contradiction is resolved if we admit that the
vector fields acquire nonzero vacuum expectation values
 $\langle \vv_x \rangle =\vv_{{\rm cl}}$,  $\langle \h_x \rangle =\h_{{\rm
cl}}$
determined by the saddle-point equations
\eqn\axi{ \h_{{\rm cl}}=\beta \hat C \vv_{{\rm cl}}= 2D\beta
\vv_{{\rm cl}},\ \  \vv_{{\rm cl}}= \ta  \h_{{\rm cl}}}
which are solved by  $     \h_{{\rm cl}}= \vv_{{\rm cl}}=0$
( strong coupling  $U(N)$-symmetric phase) or
 by  $m^2= 1- 2D\beta \ta =0$ with
    $ \h_{{\rm cl}}, \vv_{{\rm cl}} \ne 0$ ( weak coupling
 phase with spontaneously
broken $U(N)$ symmetry).
 Geometrically the weak coupling phase is characterized by
possibility of the
  the random walk to
 break into two pieces.
Then the two-point Greens function is a sum of  connected and  disconnected
parts
\eqn\gggr{G_{xy}=
\vv_{{\rm cl}}^2+\langle x|{\ta \over 1- \ta \beta \hat C}|y\rangle}
The unimodularity condition now reads
\eqn\rrr{
\vv_{{\rm cl}} ^2
 +{1 \over \beta}\int {d^Dp \over (2\pi)^D} {1\over
\sum_{\mu=1}^D(2\sin (p_{\mu}/2))^2}
=\vv_{{\rm cl}} ^2
+{\beta _c  \over \beta}=1}
Thus in the weak coupling phase there is a spontaneous magnetization
$\vv_{{\rm cl}}
=\sqrt{1-\beta_c/\beta}<1$ and  $2N-1$ Goldstone excitations
with zero mass due to the spontaneous breaking of the $U(N)$ symmetry.

In  $D=1$ dimensions the  vector model
has no dynamical degrees of freedom and is  analogous  to the
 $D=2$ gauge theory.
In terms of  the weak coupling parameter  $\lambda$ related to $\beta$ by
$2/\beta = \sinh (\lambda /2)$ we find
\eqn\pppao{\ta = \tanh {\lambda \over 2}  ,
\  \  \beta = {1 \over 2 \sinh \lambda /2}}
 and the two-point correlator is given by
\eqn\dimii{G_{xy}=e^{-(\lambda/2) |x-y|}}
This  two-point correlator can be interpreted either as the sum over all random
walks (with backtrackings) connecting the points $x$ and $y$,
 or as the only minimal path (without backtrackings)
  connecting the two points,
with a weight factor $G_1=e^{-\lambda /2}$ associated with each of its links.
In the first interpretation the mean value of $\ta$
is tuned so that
it compensates
completely the entropy due to backtrackings.
In the case of the gauge theory
the analog of the backtrackings
are the folds of  the world sheet of the string.
We shall see that the $D=2$ gauge theory can be formulated in terms of
minimal nonfolding surfaces.

\newsec{The $U(N)$ gauge theory}

In the Wilson-Polyakov formulation of the $U(N)$
lattice theory \wils\
the independent fluctuating variable is a unitary matrix
${\bf U_{\l}}=\{(U_{\l})_i^j, i,j=1,...,N\}$
\eqn\unitt{{\bf U}_{\l}^{\dag}{\bf U_{\l}}={\bf I}} associated with the
oriented links $\l = <xy>$ of the hypercubical lattice\foot{We will use the
notation $\l^{-1}=<yx>$ for the link obtained from $\l = <xy>$
by reversing the orientation.
More generally, if  $\Gamma_1$ and $  \Gamma_2$
  are two oriented lines, the product  $\Gamma_1 \Gamma_2$ is defined,
in case it exists, as the
line obtained by identifying the end of   $\Gamma_1$ with the beginning
of $ \Gamma_2$.}.
The measure and the interaction are invariant with respect to local
gauge transformations
\eqn\gggee{{\bf U}_{<xy>} \to {\bf U}_x  \U_{<xy>}\U_{y}^{\dag}}
The amplitude  $\U(\C)$ associated with the parallel
 transport along the loop
$\C=\{\l_1\l_2 ... \l_n\}$ is given by the product of the link variables
\eqn\spomn{\U(\C)= \U_{\l_1} \U_{\l_2}...\U_{\l_n}}
The action in the gauge theory is a functional on the loop fields \spomn .
The simplest nontrivial
 loops  $\C$ are associated with the elementary squares
of the hypercubical lattice which are also calles
 $plaquettes$.
 In what follows we will denote by $\U_p$
the ordered product of the  4 link variables along the boundary of the
plaquette $p$.
Let
$S_{\lambda}(\U) $ be the  one-plaquette action.
It contains as a parameter
the coupling constant  $\lambda$  of the lattice gauge theory.
The  partition function is defined by the integral over all link variables
\eqn\discr{
Z =e^{N^2 \CF}= \int _{\l \in \CS} [d\U_{\l}]
\prod_{p }e^{S_{\lambda}(\U_ p)}}
where
\eqn\measr{[d\U]= d\U d\U^{\dag} \delta(\U^{\dag}\U-\I)
\equiv  \prod_{i,j=1}^{N} dU^j_id (U^{\dag})^j_i \delta \Big(
\sum _k  (U^{\dag})_i^k U_k^j -\delta _i^j\Big)
}
 is the invariant measure on the group $U(N)$.
By
 $\CF$ we denoted the  free energy per degree of freedom.

The one-plaquette action is subjected to
the following two requirements.
First, it should be a  real function
defined on the
conjugacy classes of the group $U(N)$.
This means that it can be Fourier-expanded in
 the characters of the irreducible
representations of the group $U(N)$
\eqn\chpo{e^{N^2S_{\lambda}(\U)}= \sum_{-\infty
\le n_1\le ...\le n_N\le \infty}
\chi_{\vec n}(\I)\chi_{\vec n}(\U) e^{N^2S_{\lambda}^{\vec n} }}
The character of the representation with signature $\vec n
=\{ n_1,...,n_N\}$ depends on
the eigenvalues $u_1,...,u_N$ of the unitary matrix $\U$ as
\eqn\chhr{\chi_{\vec n} (\U)= {\det_{ik} (u_i^{n_k+k-1}) \over \det _{ik}
(u^{k-1}_i) }.}
Second, we assume that the continuum limit is achieved when $\lambda \to 0$
and in this limit
\eqn\llla{S_{\lambda}(\U) \to -{1 \over 2 \lambda  } {\tr \over N} {\bf A}^2
, \ \ \  \U=e^{i {\bf A}}.}
The Fourier image of the action in this limit is proportional to
eigenvalue of   the second
Cazimir operator $\hat C_2 =- \tr (\U \p /\p \U)^2$
\eqn\rsrsr{S_{\lambda}^{\vec n} =- {\lambda \over 2N^3} C_2(\vec n ) =
-{\lambda \over 2N} \Big( \sum _{i=1}^N n_i^2
+\sum _{i<j} (n_i-n_j)\Big)}
We will  assume that eq. \rsrsr\ remains true
 also for finite values of $\lambda$.
The corresponding one-plaquette action
\eqn\hkkk{S_{\lambda}(\U)={1\over N^2 }\log
 \sum_{-\infty
\le n_1\le ...\le n_N\le \infty}
\chi_{\vec n}(\I)\chi_{\vec n}(\U) e^{ \lambda  C_2(\vec n)/(2N)}
={1\over N^2 }\log \langle \I | e^{\lambda  \hat C_2/(2N)}|\U\rangle
}
 is known as {\it heat kernel action}
\ref\notrn{N.S. Manton, \pl 96B (1980) 328;
P. Menotti and E. Onofri, \np B190 (1984) 288;
C.B. Lang, P. Salomonson and B.S. Skagerstam, \pl 107B (1981) 211,
\np B190 (1981) 337}.
The exponential of this action
 is a solution of the  heat kernel equation on the group manifold
\eqn\hhhhk{ \big[ 2{\p \over \p\lambda}
+ {\tr \over N}(\U\p / \p \U)^2\big]e^{N^2S_{\lambda}^{(\U)}}=0}
The heat kernel action has the nice property of reproducing itself
after  group integration   \ref\ros{P. Rossi, {\it Ann. Phys.}
132 (1981) 463}
\eqn\reppo{ \int [d\U] e^{N^2S_{\lambda_1}(\U_1\U)}
 e^{N^2S_{\lambda_2}(\U_2\U^{-1})}
= e^{N^2S_{\lambda_1 +\lambda_2}(\U_1 \U_2).}}
Therefore in two dimensions the heat kernel action has trivial scaling.

Originally the lattice gauge theory was defined
in terms of the action
 \eqn\www{ S_{\lambda}^W (\U) = {1 \over 2 \lambda} {\tr
\over N}( \U +\U^{\dag})}
 known as the {\it  Wilson action}.
However this is not the best choice in our problem
since its  Fourier image is given by two different analytic expressions
in the limit $N\to \infty$, depending on the value of the coupling
constant.
For example, for the fundamental representation $\vec n_f =
[0,0,...,0,1]$
\eqn\kkkdk{S_{\lambda}^{W \ \vec n_f}=
\cases{
{1 \over 2\lambda } , &if $\lambda >1$; \cr
1-{1 \over 2}\lambda , & if   $\lambda <1$ \cr}}
The nonanalyticity at the point $\lambda =1$  known as Gross-Witten
phase transition  \ref\gw{D. Gross and E. Witten, \pr D21 (1980) 446}
is due to the fact that the integral for the
 inverse Fourier transform
\eqn\hahaha{e^{N^2S_{\lambda}^{W \ \vec n}}={1
 \over \chi_{\vec n}(\I)} \int [d\U] e^{N^2S_{\lambda}^W
(\U)} \chi _{\vec n}(\U)}
 is saturated by
the vicinity of a saddle point and the saddle-point solution can have two
analytic branches.
In the strong coupling domain $\lambda >1$  the eigenvalues
of the unitary matrix $\U$ are distributed all along the unit circle,
and  in the weak coupling domain $\lambda <1$  the density of
the phases of  the eigenvalues
is supported by an interval $[-\phi , \phi]$ with $\phi < \pi$.

\subsec{The gauge theory in terms of nonrestricted complex matrices}

Now let us convert this model into a theory with flat measure and
regular interaction using the Laplace transform of the measure in the
fields  as in the
$U(N)$ vector model. This has been actually done in the mean field
analysis of lattice gauge theories
 \jb .
We insert the exponential representation of the $\delta$-function
\eqn\dddd{\delta (U^j_i, V^j_i)= ( {1 \over 2\pi i})^2
\int_{-i\infty}^{i \infty}dH^i_j d H^i_j \ e^{ N\bar H^i_j(U^j_i-V^j_i)+
N H^i_j(\bar U^j_i-\bar V^j_i)}}
in the definition of the partition function for every matrix element
$(U_{\l})^j_i$ of the gauge fields. Here $\V_{\l}$ and $\H_{\l}$
are unrestricted complex matrices on which the integration is
performed with a flat measure
\eqn\mmmemem{d\V=\prod_{i,j=1}^N dV^i_jd\bar V^i_j,
\ d\H=\prod_{i,j=1}^N dH^i_jd\bar H^i_j}
For each oriented link $\l = <xy>$ only one pair
 of these fields is introduced
and the convention
\eqn\yty{
\V_{<yx>}=\V_{\l^{-1}}= \V^{\dag}_{\l}=\V^{\dag}_{<xy>},\ \ \
\H_{<yx>}=\H_{\l^{-1}}= \H^{\dag}_{\l}=\H^{\dag}_{<xy>}}
is assumed.
Denoting by
\eqn\onlin{e^{N^2F(\H)}=\int [d \U] e^{N\tr (\H^{\dag}\U+\H\U^{\dag})}}
the one-link mean-field integral
we write the partition function \discr\ in the form
 \eqn\oi {\eqalign{
\CZ=e^{N^2\CF}&= \int \prod _{\l }\Big(
d\H_{\l}d\V_{\l}
e^{-N\tr (\H^{\dag}_{\l}\V_{\l} +\H_{\l}\V^{\dag}_{\l})}\Big)\cr
&\prod_{\l} e^{N^2
F(\H_{\l})} \prod _{p}e^{N^2 S(\V_ p)}\cr}}
where $\V_ p$ denotes the ordered product of link variables along the
boundary $\p p$  of the elementary square $p$
\eqn\cads{\V_ p=\prod_{\l \in \p p} \V_{\l}.}
The measure is now flat and the integrand is regular.
The constraint \unitt\ is satisfied in the sense of mean values
\eqn\unvt{\langle (\V^{\dag}_{\l}\V_{\l}\ -\I)  ...\rangle =0}

 Assuming that the saddle point for
the integral  \oi\ is  the trivial field  $\H=\V=0$,
we can express the free energy as a sum of vacuum
Feynman diagrams. The functional integral can be calculated by expanding
the exponentials and applying all Wick contractions between the
$\H$ and $\V$ fields
\eqn\connt{\langle  (H_{\l})^i_j (V^{\dag }_{\l})^k_l \rangle =
\langle  (H^{\dag}_{\l})^i_j (V_{\l})^k_l \rangle
={1 \over N}\delta^i_l \delta^{k}_j}

\subsec{The $U(N)$ gauge theory \`a la Weingarten}

In order to give topological meaning of the  $1/N$ expansion
we need to expand the two potentials $F$ and $S_{\lambda}$
in terms of the moments
\eqn\mmmi{\a_n={ \tr \over N} (\H^{\dag}\H)^n,}
\eqn\mimmi{ \beta _n=
{ \tr \over N} \V^n , \  \beta _{-n}={ \tr \over N}
  \V^{\dag n}
;\ \ \ \  \ n=1,2,... }
of the matrix fields.
The coefficients in the  the expansions
\eqn\odof{ F[\a]=\sum_{n=1}^{\infty}
\sum_{k_1,...,k_n \ge 1}
f_{[k_1,...,k_n]}{\a_{k_1}
...\a_{k_n} \over n!}}
\eqn\asda{N^2 S_{\lambda}[\beta]=N^2 \sum_{n=1}^{\infty}
 \sum_{k_1,...,k_n \ne 0}
s_{[k_1...k_n]} {\beta _{k_1}...\beta _{k_n}
\over n!}}
can  be found from the corresponding Ward identities (Appendix A).

The lowest-order coefficients are
\eqn\exx{\eqalign{
f_{[1]}=1 , \  f_{[2]}=-1 \ f_{[1,1]}=1, \  f_{[3]} =2 , \ f_{[4]}=-5 \cr
s_{[1]}=s_{[-1]}=e^{-\lambda /2}, \ s_{[2]}=s_{[-2]}
=(\cosh  {\lambda \over N})
-N\sinh {\lambda \over N})e^{-\lambda}, \cr
\ s_{[1,1]}=s_{[-1,-1]}=
\big[-N\sinh {\lambda \over N}+ 2\big( N \sinh  {\lambda \over 2N}\big)^2
\big] e^{-\lambda},
s_{[1, -1]}=-e^{-\lambda}. \cr}}

The interpretation of the functional integral as a sum over surfaces is
possible if the action is linear in the traces of the matrix fields.
This is achieved by the integral transformation
\eqn\eff{e^{N^2F(\H)}=
\int
\prod_{n=1}^{\infty}
(d\a_n d\ta_n
\ e^{
N^2[ { \ta_n \over n} {\tr\over N}  (\H^{\dag}\H)^n  -
{\ta_n\a_n \over n}] }) \ \ e^{ N^2 F[\a]}}
By means of another system of parameters
we represent the exponential of the plaquette  action
 in the form
\eqn\sssos{
e^{N^2S( \V)}=
\int
\prod_{n=1}^{\infty}
(d\beta _n d\tilde \beta _n d\beta _{-n} d\tilde \beta _{-n}
e^{N^2[ {\tilde \beta _n \over n} ({\tr\over N} \V^n - \beta _n )
+ {\tilde \beta _{-n}\over n}( {\tr\over N} \V^{\dag n}
  -\beta _{-n})]}
) e^{N^2  S[\beta ]}
}

We therefore introduce at each link $\l$ and at each elementary square  $p$
a set of auxiliary loop variables
coupled to the moments of the matrix fields
\eqn\aaapa{\a_n (\l), \ta_n(\l); \  \
 n=1,2,...}
\eqn\vvvpv{\beta_{n}(p),\tilde \beta _n(p);\ \ n=\pm 1, \pm 2, ...}
We call them {\it puncture operators}
  by analogy with the matrix models for noncritical string theories.
After that we are able to  represent the integral \oi\ as
  a theory of Weingarten
type
 described by the partition function
\eqn\lst{\eqalign{
\CZ =e^{N^2\CF}&=
 \Bigg\langle  \int \prod _{\l }
d\H_{\l}d\V_{\l}
e^{-N\tr (\H^{\dag}_{\l}\V_{\l} +\H_{\l}\V^{\dag}_{\l})}\cr
&\prod _{\l ; n>0}e^{{N \over n} \ta_n(\l) \tr (\H_{\l}^{\dag }
\H_{\l})^n }
\prod _{p; n>0}e^{{N \over n}[ \tilde \beta _n(p) \tr \V_p^n
 +\tilde \beta _{-n}
 \tr \V^{\dag n}_p]}  \Bigg\rangle_{\ta , \tilde \beta } \cr}
}
where the Boltzmann weights of the surfaces are themselves
quantum fields and the average with respect to them is defined by
\eqn\rrrea{\eqalign{
\langle  \cdot  \rangle
_{\ta  , \tilde \beta } &=\int \prod _{\l \in \CL;n>0}
 d\a_n(\l)d\a_n(\l)
e^{-N^2 {\ta_n(\l)\a_n(\l)\over n} +F[\a(\l)]}\cr
&\prod_{p;n \ne 0} d\tilde \beta _n(p)d\beta _n(p)
e^{-N^2{\tilde \beta _n(p) \beta _n(p) \over n}+S[\beta (p)]} \cdot \cr
  }}

The integration over the matrix fields in \lst\ will produce
an effective action which will be interpreted in the next section
as a sum over closed connected surfaces on the lattice.
The leading term in this action is proportional to $N^2$.
Therefore, in the limit $N\to\infty$ the integral over the scalar
fields \aaapa\ and \vvvpv\ is saturated by a saddle point and
these fields freeze at their vacuum expectation  values
\eqn\clls{\langle \ta_n(\l)\rangle=\ta_n , \ \ \
\langle \tilde \beta _n (p) \rangle  =
\langle \tilde \beta _{-n} (p) \rangle  = \tilde \beta _n}
The fluctuations of the scalar fields around their mean values
are of order $1/N$ and  their correlations will appear  in the
nonleading terms  of the $1/N$ expansion.

If we are interested only in the large $N$ limit, the fields
\aaapa\ and \vvvpv\ can be replaced with their vacuum expectation
values even before the integration over the matrix fields.
We therefore find the following expression for the free energy of
the $U(\infty)$ gauge theory
\eqn\aiaiai{\eqalign{
\CF =\lim_{N\to\infty}{1 \over N^2}\log &
\Bigg(\int
  \prod _{\l }
d\H_{\l}d\V_{\l}
e^{-N\tr (\H^{\dag}_{\l}\V_{\l} +\H_{\l}\V^{\dag}_{\l})} \cr
& \prod _{\l ; n>0}e^{{N \over n} \ta_n\tr (\H_{\l}^{\dag }
\H_{\l})^n }
\prod _{p; n>0}e^{{N \over n} \tilde \beta _n ( \tr \V_p^n
 +
 \tr \V^{\dag n}_p) } \Bigg)\cr}
}
The parameters  $\ta_n$ are determined by the condition of unitarity
\unvt\ and $\tilde \beta _n$ can be considered as coupling
new constants defining the plaquette action.

\newsec{Lattice strings (random surfaces)}

\subsec{Feynman rules for the matrix fields }
By  applying the rules of the Feynman-'t Hooft diagram expansion,
 the partition function of the gauge model \oi\ can be reformulated  as
the partition function of a gas  of lattice surfaces.
An important feature of  these surfaces is that they can
have local singularities of the
curvature not only at the sites, as it is the case in the original
Weingarten model, but also at the 1-cells (links) and 2-cells
(plaquettes) of the lattice.
  These surfaces are
composed of  elements with the topology of a disk,
 glued together by  identifying pairwise pieces of
 edges with opposite orientations  along the links of the lattice
 (this is the meaning of the contractions
\connt ). There are two types of   surface elements
 associated with the
 $\ta$ and $\tv$ vertices.

\leftline{{\it a) link-vertices}}
The vertex  $N\ta_n(\l)\tr(\H^{\dag}_{\l}\H_{\l})^n$
  will be represented graphically
by a punctured disk with
 boundary going along the loop
  $(\l \l^{-1})^{n}$. The $\H$-matrices are associated with the $2n$ edges
and the operator  $\tilde \a_n$ - with the puncture (\fig\faui{
Three different ways of drowing the elementary
 surfaces associated with link-vertices.
The edges are deformed  for eye's convenience. }).
This surface element has zero area and
Gaussian curvature   $2\pi (n-1)$ which we
associate with  the puncture.

\leftline{{\it b) plaquette-vertices}}
The  vertex $ N
\tv_{n}(p) \tr \V^{n}_p$ corresponds to
  a punctured  disk with area $n$
covering $n$ times the plaquette
 $p$ and having as a boundary the loop
 $(\p p)^n$
 going $n$ times around $p$
 (\fig\fauii{Elementary surfaces
associated with plaquette-vertices}).
We will refer to this surface element as $n$-plaquette,
or {\it multiplaquette} of order $n$.
The $\V$-matrices are associated with the edges of the boundary
and the $\tv$-operator - with the puncture. As before, the
curvature $2\pi (n-1)$ will be associated with the puncture.

The vertex  $ N
\tv_{-n}(p) \tr \V^{\dag n}_p$ corresponds to the
same surface but with the opposite orientation.
Note that the $\tv_n$-operator changes to $\tv_{-n}$ after changing the
orientation. This means that the puncture is not just a point, but
rather an oriented infinitesimal  loop.

Expanding the exponential in a sum of monomials, we can assign to each
monomial a set of multiplaquettes of different orders
covering the 2-cells (plaquettes)
and a set of link elements associated with the  1-cells (links)
of the lattice.
Then we apply all possible contractions
  \connt\ between $\V$ and $\H$ matrices.
The geometrical meaning of these contractions is identifying oppositely
oriented edges of link and plaquette surface elements.
 The result is a collection of closed surfaces composed from multiplaquettes
glued together by means of cyclic contractions (the link vertices).
A piece of such surface is
shown in \fig\ftre{Two ways of drawing a piece of a  surface obtained by
identifying edges of surface elements. The link-vertex of order 2
creates a curvature $2\pi$.}.
The weight of each surface is the
 product of the  $\ta$ and $\tv$ operators  associated with the
punctures.

Alternatively, one can imagine these surfaces as made only of
$n$-plaquettes glued together along their half-edges. Such was the
interpretation given in ref. \ivn .

The dependence on the number of colors $N$ is, as in all $N\times N$
matrix models,  through a factor $N^{\chi}$
where $\chi$ is the total Euler characteristics.

The sum over all
surface  configurations
  is given by  $e^{\CF_0[\ta,\tv]}$
where  $\CF_0[\ta,\tv]$
 is the sum over all
connected   surfaces\foot{The trivial exponentiation  is the main
advantage of the present method compared with the character expansion
 used in \dg , \dgwt .}
\eqn\frener{\CF_0[\ta,\tv] =
\sum_{\CS: \p \CS =0} N^{\chi (\CS)}
\  \Omega[\ta,\tv]
}
In the last formula  $\chi(\CS)$ is the Euler characteristics
 of the surface
and the $\Omega$-factor is the product of all
 $\ta ,\tv $-operators associated with the elementary disks from which the
surface $\CS$ is composed.
The  partition function of the gauge theory  is
obtained  as  the expectation value
  \eqn\absi{
e^{\CF} =
\langle  e^{ \CF_0[\ta,\tv]}
\rangle}

In this way the original gauge theory is formulated as a model of
lattice random
surfaces whose Boltzmann weights are products of
local gauge-invariant puncture operators $\ta_n, \tv_n$.
It follows from \rrrea\ that the  interaction of these operators, mediated
by the fields  $ \a_n, \v_n$,   is
of order $1/N^2$. The vertices of this interaction are the coefficients
in   the expansions \odof\ and \asda .
They  can be interpreted as
microscopic
surfaces with several punctures (\fig\fint{Vertices
for the contact interactions between surfaces}).

The classical values of the puncture operators are determined by the
 saddle point equations
\eqn\exxx{{1 \over n} \a_n={\p \CF _0 \over \p  \ta_n}
={1 \over n}\langle (\H^{\dag}\H)^n
\rangle , \
{1\over  n} \v_{n}={\p \CF_0 \over \p \tv_n}={1\over n}
\langle \V_p^n\rangle ;\
{1\over n}\ta_n={\p F \over \p \a_n},\ \
{1\over n}\tv_n={\p S_{\lambda} \over \p \v_n}}
In the limit $N \to \infty$
the free energy is given by the sum of
noninteracting random surfaces
\eqn\clfe{\CF=\CF_0(\ta, \tv)+
\sum_{\l}F[\a]+
\sum_{p}S_{\lambda}[
\v]- \sum_{\l ,n\ge 1}
 {1\over n}\a_n\a_n- \sum _{p, n\ne 0} {1\over n}\tv_n\v_n
}

Applying the same arguments to the Wilson loop average
$W(\C)= \langle {\tr \over N} \V(\C)\rangle$
we find in the limiy $N\to\infty$
a representation of $W(\C)$ as the sum over all planar surfaces
bounded by the contour $\C$.
\eqn\kkksk{
W(\C)=  \sum_{\CS:\p \CS =\C}
\Omega [\CS]}
The Boltzmann weight of a surface  is the product of
the mean values \clls\
associated with the punctures.
The $\Omega$-factor of the surface $\CS$ containing
  $\CN^{(p)}_n$
$n$-plaquettes, $n=1,2,...$,
  and  $\CN^{(\l)}_n$
 cyclic contractions of order $n$ ,
$n=1,2,...$, is given by
\eqn\oomomo{
\Omega [\CS]=
\prod_{n=1}^{\infty}
 (\tb_n)^{\CN^{(p)}_n } (\ta_n)^{\CN^{(\l)}_n}}

\subsec{Irreducible surfaces (a miracle)}

Unlike the random walk, the two-dimensional surface can exist in
configurations with very uneven intrinsic geometry.
The typical syngularity is a ``neck'' representing an
intermediate closed string state of small length
 connecting a ``baby universe''
with the main body of the surface.
The most singular configurations characterized by  necks
occupying a single link
(\fig\bubb{A reducible surface})
 can be readily removed from
the path integral by absorbing their contribution
in the Boltzmann weights of the  punctures.
  Namely, the sum over all reducible
surfaces containing ``baby universes'' renormalizing the link vertices
 can be taken into account by
 adjusting the
classical  values $\ta_n$  the puncture operators located at links.

In general, the
 classical fields  $\ta_n,\tv_n$
 are complicated functions of the coupling
constant $\lambda$.
Miraculously, the new Boltzmann weights
  become easily
calculable and do not depend neither on the dimension $D$ nor on the
choice of the one plaquette action!
This is why we will restrict the sum $\CF_0$  over all
connected  surfaces to the sum $\CF_I$ of
irreducible ones which will render the
 random surface {\it Ansatz} much simpler than it its original version.

\smallskip
\noindent {{\it Definition:}}
\smallskip
{\parindent =4em
\narrower
\noindent
A surface $\CS$ immersed  in a lattic $\CL$
is {{\it reducible}} with respect to given link
$\l  \in \CL$  if it splits  into two
or more
 disconnected pieces after being cut along
this link.
A surface which is
not  reducible with respect to  any  $\l \in \CL$ is
 is called {\it irreducible}.

\par}

If we restrict the sum over surfaces to the irreducible ones, the
free energy is obtained by the same formulae
 \exxx\ and \clfe, with $\CF_{0}$ replaced by $\CF_I$.
In what follows  by sum over surfaces we   will understand
a sum over irreducible surfaces.
This allows to simplify drastically  the classical equations of motion
\exxx\
and solve them exactly.
It turns out that, with this replacement, the expectation
values of the puncture operators $\ta_n$
coincide with the coefficients $f_{[n]}  $
in the expansion\odof .

\subsec{ Evaluation of the link-vertices}

Instead of solving directly \exxx\ we would like to present
 a short-cut derivation
using the hidden unitarity of the $\V$-field.

Consider  the Wilson loop average for the
 closed contour $\C=(\l_1\l_2...)
$ in  the lattice $\CL$. It is given by eq. \kkksk\
in which the sum is restricted to irreducible surfaces.

We will exploit the unitarity condition $\V\V^{\dag} =\I$,
applied to the loop amplitude.
It means that the Wilson average $W(\C)$
will not change if a backtracking piece
$\l \l^{-1}$ is added to the contour $\C$
\eqn\unilop{W(\C \l\l^{-1})=W(\C)}

The sum over surfaces spanning the loop $\C \l\l^{-1}$ can be divided into
two pieces
\eqn\lpol{W(\C\l\l^{-1})=W(\C)
W(\l\l^{-1}) + W_{{\rm conn}}(\C \l\l^{-1})}
The first term is the sum over all  surfaces made of two disconnected
parts spanning the loops $\C$ and $\l\l^{-1}$. The  second term
contains the rest.
The constraint \unilop\ is satisfied for all loops if $W(\l\l^{-1})=1$
and $W_{{\rm conn}}(\C \l\l^{-1})=0$.
here is only one irreducible surface spanning the loop $\l\l^{-1}$;
it contains a single contraction  $\ta_1(\l)$ between  $\l$ and $\l^{-1}$.
Therefore $\ta _1 =1$.

Now consider a surface contributing to the second term $W_I$.
The
links  $\l$ and $\l^{-1}$  may be connected to the rest of the surface
 by means of  the same link vertex  or  by  two different
link vertices.
  The condition that their total contribution is zero
is
\eqn\rscx{\ta_n + \sum _{k=1}^{n-1} \ta_k  \ta_{n-k}
=0, \ \ n=2,3,...}
Eq. \rscx\ is identical to the loop equation  satisfied by the
coefficients $f_n$ in the expansion \odof , whose unique solution
is given  by the Catalan numbers (see  Appendix A)
\eqn\iihy{ \ta_{n}=f_{[n]}=(-)^{(n-1)}{(2n-2)! \over n! (n-1)!}}

Since the evaluation of the link vertices
is the most important point in the whole construction, we
give a  second
 derivation which
is based only on the
 the sum over surfaces for the
trivial Wilson loops
\eqn\trivv{W[(\l \l^{-1})^n] = \langle {\tr \over N} (\V_{\l}\V^{\dag}_{\l}
)^n \rangle =1, \ n=1,2,... }
For each of these Wilson loops the sum over irreducible surfaces
contains only finite number of terms, namely,
the link-vertices contracting directly the edges of the loop
$\l\l^{-1}$. For example,
\eqn\axa{W(\l\l^{-1})=\ta_1 ,W[(\l\l^{-1})^2]=2\ta_1 ^2+\ta_2, ...}
Introducing the generating functions
\eqn\genn{w(t)= \sum_{n=0}^{\infty}t^n W[(\l \l^{-1})^n]
={1 \over 1-t}, \ \ f(t)=1+\sum_{n=1}^{\infty}t^n (\ta_n)^n}
we easily find the relation \ivn
\eqn\stary{w(t)=f[tw^2(t)]}
which is solved by the generating function of  the Catalan numbers
\eqn\fff{f(t)={1+\sqrt{1+4t} \over 2}.}

In this way, unlike the $U(N)$ vector model, the classical
values of the auxiliary fields
 {\it do not depend on the dimension of the space-time}.
The expectation value  $\ta$ in the vector model is determined by the
long wave excitations of the random walk.
Here in the gauge theory, the expectation values of the puncture operators
$\ta_n$ are determined in purely local way, as it is clear from
their derivation, and coincide with the coefficients $f_{[n]}$ in the
expansion \odof\ of the potential $F(\a)$.
One way to explain this difference is the local character
of  the $U(N)$ invariance in the case of the gauge theory.

On the contrary, the weights $\tv_n$ of the
multiplaquettes  will depend on the dimension as
well as on  the choice of the one-plaquette action.
They coincide with the coefficients $s_{[n]}$ in the expansion \asda\
only in the trivial case $D=2$.

If we are interested only in the large $N$ limit,
it is more convenient
 not to try to calculate them by solving the equations of motion
but just to take them as independent coupling constants.

We have found that after restricting the sum over surfaces
to the irreducible ones, the  expectation values of the puncture operators
at the links are the same as the bare ones. That is, the renormalization of
the bare link vertices $f_{[n]}$ due to tadpole diagrams
is neatly compensated by the contribution of the reducible surfaces.
    This is true also for
the nonplanar vertices responsible for the contact interactions
of surfaces. The equations for the renormalized link vertices
obtained from the surface representation of the multiloop correlators
$\langle {\tr \over N}(\H_{\l}\H^{\dag}_{\l})^{k_1}
{\tr \over N}(\H_{\l}\H^{\dag}_{\l})^{k_2}...\rangle =1$ are equivalent
to the loop equations for the bare vertices considered in Appendix A.

\subsec{Loop equations}
In ref. \ivn\ we proved that
the  loop equations
in  the Wilson lattice gauge theory
\ref\migg{
A.A. Migdal, $unpublished$ (1978);
D. F\"orster, \pl 87B (1979) 87; T. Eguchi, \pl 87B (1979) 91}
 are satisfied
by the
the sum over  surfaces for the Wilson loop average.
Let us  only write here  the general formula which is derived in the same
fashion.
For any link $\l \in \Gamma$
\eqn\lopsr{\sum_{n=1}^{\infty} \sum _{p: \p p \ni \l} \tilde \beta_n
[(W(\C (\p p)^n)-W(\C(\p p)^{-n})]=
\sum _{\l' \in \C}W(\C _{\l\l'})W(\C_{\l'\l})
[\delta(\l , \l')- \delta(\l^{-1} , \l ')]}
The sum on the l.h.s. goes over  the $2(D-1)$ plaquettes
adjacent with the link $\l$ and the closed loops $\C_{\l \l'},
\C_{\l' \l}$ in the contact term
are obtained by cutting the links $\l$ and $\l'$ and
reconnecting them in the  other possible way.

Eq. \lopsr\ is the lattice version of the loop equation \mme\ in the
continuum theory.
\newsec{The trivial $D=2$ gauge theory
as a nontrivial model of random surfaces}

The above construction can be easily checked in
 two dimensions where one can perform explicitly the sum over surfaces.
The motion of the string is restricted by  kinematics to the longitudinal
vibrations and one can show that the
contribution of the longitudinal modes of the string is zero, i.e.
the surface maken no folds.
 The only singularities of a nonfolding surface
are those  associated with the punctures. A singularity containing
curvature $2\pi (n-1)$  can be interpreted
as branch points of order $n$. Therefore the Wilson loop average  in two
dimensions can be expressed as a sum over all surfaces with
minimal area  and eventually  having branch points,
bounded by the loop. This form of the random surface {Ansatz} in
two dimensions was
anticipated in \kk\ and established in \ivn .

\subsec{The contribution of the surfaces with folds is zero}
The weights of the branch points, as they were determined
by eq. \rscx , provide a mechanism of suppressing the backtracking
motion   of the strings or, which is the same, world surfaces having folds.
The  surfaces with folds have both positive and negative weights and
their total contribution to the string path integral is zero.

We have checked that in many particular cases but the general proof is missing.
An indirect proof might be constructed using the
property \reppo\ of the heat kernel action,  which allows to
delete parts of the lattice $\CL$ without changing  the sum over surfaces.

It is perhaps instructive to give one example. Consider the
surface in \fig\ffold{ A piece of surface having a fold}.
It  covers three times the interior of the nonselfintersecting
loop $\C$ and once the rest of the lattice.
The change of the orientation of the world surface occurs along the
loop.
There are two specific points on the loop $\C$ at which the curvature
has a conical singularity; they can be thought of as the points
where the fold is created and annihilated.
Each of these points can occur either at a site or at a link (in the
last case it coincides with the branch point of order two of a link-vertex).
Let us evaluate the total contribution of all irreducible  surfaces
 distinguished by the positions of the two singular points.
Denoting by $n_0$ and $n_1 (=n_0)$
 the number of sites and links along the loop $\C$,
and remembering that a singular point located at a link has to be taken
with a weight $\a_2=
f_{[2]}=-1$, we find that the contribution of these surfaces
is proportional to
\eqn\ffafa{[{n_0(n_0-1) \over 2}+f_{[2]}n_0n_1+f_{[2]}^2{n_1(n_1-1)\over 2}]
-[n_1 +2 f_{[2]}n_1] =0}
The second term on the left hand side contains the contribution of the
reducible surfaces which must subtracted.
A reducible surface arises when the two points are located at the extremities
of the same link ($n_1$ configurations) or when one  of the points is
a   branch point and the other is located at one of the extremities of the
same link ($2n_1$ configurations).

\subsec{ Evaluation of the plaquette-vertices}

Now we are ready to  evaluate the expectation values of the fields $\tv_n$.
For this purpose we
consider  the sum over surfaces   for
the one-plaquette Wilson loops $W((\p p)^n); n=\pm 1, \pm 2, ... $
in the limit of infinite area  $A_{{\rm tot}}$
covered by  the lattice $\CL$.
In this limit
 the one-cell Wilson loops do not feel the rest of the lattice
and coincide with these for the one-plaquette model.
(This well known fact  specific to two dimensions  can be proved
using the character expansion).
These loops  can be calculated from the
loop equations (Appendix A); they  coincide with
 the coefficients $s_n$ in the expansion \asda .

First let us observe that the only surfaces contributing to the sum
are the surfaces with minimal area $n\lambda$ covering only the cell $c$.
All other surfaces have folds (this is true only for an infinite
lattice) and their contributions cancel.
Therefore the sum over surfaces
 contains only finite number of terms and can be
easily evaluated
\eqn\onplq{\eqalign{
 s_{[n]}&=W((\p p)^n)= (\ta_1 )^{4n}\tv_n \cr
&+
4 n(\ta_1)^{nl-2} (\ta_2 +(\ta_1)^2)
 \sum _{k=1}^{n-1}
    \tv_k \tv_{n-k} +...\cr}
}
Applying the  identities  \rscx \  one finds
\eqn\gaga{\tv_n = \tv_{-n}=
s_{[n]}; \ \ n=1,2,3,...}

A similar statement is true for the nonplanar plaquette vertices:
amplitudes for the contact interactions of the puncture operators $\tv_n$
are given by the higher
 coefficients in the expansion \asda .

\subsec{String representation for the Wilson loops on the infinite plane.}

Now we are able to formulate the string {\it Ansatz} for the
Wilson loop average  as the sum over all
 irreducible surfaces without folds, spanning the loop $\C$,
and having branch points of all orders. It is convenient to
write  the $\Omega$-factor in \kkksk\ as $\Omega = e^{-\lambda {\rm Area}}
\tilde \Omega$. Then  the sum over surfaces \kkksk\ takes the form
\eqn\willsl{W(\C)=
\sum_{ \p \CS =\C}\tilde  \Omega
 (\CS) e^{-{1 \over 2}\lambda {\rm Area (\CS)}}}
The factor $\tilde \Omega (\CS)$ is a product of the weights of all branched
points
of the world sheet and the area of the surface is the sum of the areas of
its elements (recall that the area of an $n$-plaquette is $n$).
A branch point of order $n$ ($n=1,2,3,...$)
associated
 with a $k$-cell ($k=0,1,2$
 for sites, links, plaquettes correspondingly)  is weighed by a factor
$\omega_{n}^{(k)}$ where
\eqn\omom{\eqalign{
\omega^{(0)}_ {[n]}
&= 1\cr
\omega^{(1)}_{[n]} &= f_{[n]}=(-)^{(n-1)}{(2n-2)! \over n! (n-1)!}\cr
\omega^{(2)}_{[n]} &= s_{[n]}e^{n\lambda/2} =\sum_{m=0}^{n-1}
\pmatrix{n\cr
m+1\cr}{n^{m-1} \over m!}(-\lambda)^m
=(1-{n(n-1)\over 2}\lambda + ...)\cr}
}
In the continuum limit $\lambda \to 0$ one can leave only the
linear term in the weights $\omega_n^{(2)}$ of the $n$-plaquettes.

\subsec{Wilson loops on the infinite plane. Examples}

Let us illustrate how the
string {\it Ansatz} \willsl \ works by  some simplest
   examples of planar loops (\fig\ffig{Examples of Wilson loops:
a) Circle, b) Flower, c) Pochhammer contour}) which have been calculated
previously using the Migdal-Makeenko loop equations \kk .
In order to simplify the notations, it is convenient to measure
the area in units of $1/\lambda$. Then the dependence on $\lambda$ will
completely desappear from the final expressions.

\leftline{$a)$ {\it Circle}}
Assume that the loop encloses $n_2$  2-cells and the  total area is
$A= n_2 \lambda$.
The sum over surfaces contains a single
 nonfolding surface spanning
the loop and
\eqn\siml{W(\C)= e^{-{1 \over 2}n_2\lambda}=e^{-{1 \over 2}A}}
This  is the famous area law for the Wilson loops in ${\rm QCD}_2$.

\leftline{$b)$ {\it Flower}}

The nonfolding surfaces spanning the loop will cover the three petals
of the flower (denoted by 1 in \ffig ) once, and the head (denoted by 2)
- twice. Therefore the total area will be always $A= A_1+2A_2$,
where $A_i$ is the area enclosed by the domain $i \ (i=1,2)$.
Each of these surfaces will have a branch point located at some site,
link or plaquette  of the overlapping area 2 (\fig\faaa{A branched covering of
the flower contour}).
Let us denote by $n_{0}, n_{1 }, n_{2 }$ the numbers
of sites, links, plaquettes  belonging to the domain 2.
Then the sum  over the positions of the branch point
  reads
\eqn\omfac{\sum \tilde  \Omega =
\sum_{k=0,1,2}n_{k}
\omega_{[2]}^{(k)} = (n_0-n_1+(1-\lambda)n_2)=1-n_2\lambda}
(Here we used the Euler relation $n_0-n_1+n_2 =1$.)
Therefore
\eqn\womg{W(C)=(1-A_2) e^{-{1 \over 2}(A_1+2A_2)}}

\leftline{$c)$ {\it  Pochhammer contour}}

This time the surfaces spanning $\C$  may have different area.
The three possible cases are shown in \fig\fffirg{Three ways of
spanning the Pochhammer contour}.
There is one surface with area $A_1+A_{\bar 2}$, one surface with area
$A_1+A_2$ and a whole class of surfaces with area $A_1+A_2+A_{\bar 2}$
which are distinguished by the position of a branch point of order $2$.
The two overlapping sheets are with opposite orientations and the
branch point can occur only along the boundary of the overlapping
region. Let $n_0$ and $n_1$ be the numbers of sites and links belonging to
this boundary. By the Euler relation, $n_0-n_1 =1-2=-1$ (the
boundary has  two cycles).
Therefore the sum over the positions of the branch point contributes a
factor $\sum _{k=0,1} n_k \omega_{[2]}^{(k)} =n_0-n_1=-1$
and  the Wilson average reads
  \eqn\ovoma{W(\C)= e^{-{1\over 2}(A_1+A_2)}+e^{-{1\over 2}(A_1+A_{\bar 2})}
-e^{-{1\over 2}(A_1+A_2+A_{\bar 2})}}

\smallskip

Finally, consider  the sum over vacuum surfaces in the large $N$ limit.
It follows from \clfe\ that the  vacuum energy is equal to the  sum over
surfaces  that, after being cut along all copies of a  given link, split
only into irreducible pieces.
On the infinite plane, the only surfaces
satisfying these  conditions are made by identifying the boundaries of two
multiplaquettes and the free energy  is given by
\eqn\ffru{\CF= (\# {\rm cells}) \sum _{n=1}^{\infty} {s_n^2\over n}}
In the continuum limit $\lambda \to 0, A_{{\rm tot}}
=\lambda \# {\rm plaquettes }$ = const,
we find, up to an infinite constant,
\eqn\okok{\CF = { A_{{\rm tot}}\over \lambda}\sum _{n=1}^{\infty}
{1 \over n}(1-n^2\lambda + ...) = {1 \over 12} A_{{\rm tot}}}

\subsec{Interactions}
The $1/N^2$ corrections appeaar due to surfaces with higher topology as
well as due to the contact interactions between the punctures.
In two dimensions the interaction vertices are given by the coefficients
in the expansions \odof\ and \asda .

Let $\omega ^{(k)}_{[n_1n_2...n_m]}$ be the interaction amplitude for
the   puncture operators  of orders $n_1,n_2,...,n_m$ located at a $k$-cell
($k=1,2$).
Then, in two dimensions,
\eqn\jajaj{\omega^{(1)}_{[1,1]}=f_{[1,1]}; \ \ \omega ^{(2)}_{[1,1]}=s_{[1,1]}}

The simplest example where the interactions appear is the two-loop
correlator $W(\C_1,\C_2)$.
Let us take the simplest case of non-selfintersecting loops (\fig\fauu{
A configuration of two loops}).
Then the  answer depend on whether the two loops have equal or opposite
orientations.

\leftline{{\it I. Equally oriented contours }}

$(i)$  All cylindric surfaces bounded by the two contours have
the same area,  one cut
and  two branch
points which can appear at the sites, links and cells of the overlapping
domain 2. The sum over surfaces is a sum over the positions of these
two branch points.
The configurations leading to a reducible surface (i.e., when the two
branch points can be connected by a link) are excluded from the sum.
 Let $n_0,n_1, n_2 $ be the number of sites, links and cells of
this domain.
Then the sum over the $\Omega$-factors yields
\eqn\haidebe{\eqalign{&
{n_0(n_0-1)-n_1 \over 2} \omega _{[2]}^{(0)} \omega _{[2]}^{(0)}
+ (n_0n_1-2n_1) \omega _{[2]}^{(0)}\omega _{[2]}^{(1)} \cr
&+
n_0n_2 \omega _{[2]}^{(0)}\omega _{[2]}^{(2)}+
 {1 \over 2}n_1(n_1-1)\omega _{[2]}^{(1)} \omega _{[2]}^{(1)} \cr
& +
n_1n_2 \omega _{[2]}^{(1)}\omega _{[2]}^{(2)}+
 {n_2(n_2-1) \over 2}\omega _{[2]}^{(2)}\omega _{[2]}^{(2)}\cr
&= {1\over 2}(n_0-n_1+n_2
-2\lambda n_2 )(n_0-n_1+n_2-1) +{1 \over 2}\lambda^2 n_2(n_2-1)\cr
&={1 \over 2}\lambda^2 n_2(n_2-1)\cr}
}

$(ii)$ The contact interactions due to the interaction of the
puncture operators   lead to a second term which is the sum over
surfaces representing two discs connected with a microscopic  tube.
Their contribution is
\eqn\ipli{n_2\omega^{(2)}_{[2]}=n_2 (-\lambda +{1 \over 2}\lambda ^2)}
The sum over the two terms gives
\eqn\woas{W(\C_1,\C_2)= e^{-{1\over 2}(A_1+2A_2)}(-A_2+{1\over 2}A_2^2)}

\leftline{{\it II. Oppositely  oriented contours}}

$(i)$  Besides the minimal cylindric surface with area $A_1$ there
is a whole class of cylindric surfaces with area $A_1+A_2$ (see ref.
\ivn ).
The contribution of these surfaces is (\ivn , Appendix B)
\eqn\iaiaia{e^{-{1 \over 2}A_1}+ (n_2 -1)e^{-{1 \over 2}A_1 +A_2}
 }

$(ii)$
The contact interactions
produce a term
\eqn\ttrtr{-n_2e^{-{1\over 2}A_1+A_2}}
\eqn\avac{W(\C_1,\C_2)= e^{-{1\over 2}(A_1-A_2)}- e^{-{1\over 2}(A_1+A_2)}}
so that the total sum depends only on the two areas
\eqn\ttotl{W(\C)=-e^{-{1\over 2}A_1+A_2}+
e^{-{1 \over 2}A_1}}

\subsec{Compact lattices}
For a compact lattice $\CL$  with the topology of a sphere and area
$A_{{\rm tot}}$
the sum over surfaces becomes
much more involved.
 It will contain a
 sum over spherical surfaces
 representing  multiple branched coverings of $\CL$.
There are infinitely many
 surfaces contributing to the elementary Wilson loops. For example,
\eqn\wlplo{W_1 =  e^{-{1 \over 2}\lambda}+e^{-{1\over 2}(
A_{{\rm tot}})}+...}
Therefore, we can either  modify the weights of the irreducible
surfaces, or use the same weights as for the infinite lattice but allow
contact interactions
due to
 microscopic
tubes, trousers, etc., connecting two, three, etc. multiplaquettes.
The amplitudes of these contact interactions are given by the
coefficients in the expansion  \asda .

We are not going to discuss this case in details since there is
nothing conceptually new.
As a single example we will calculate the two leading orders of the free
energy due to the surfaces defining
simple and double coverings of $\CL$.

Let $n_i$ be the number of $i$-cells of the lattice $\CL$ ($i=0,1,2$).
The free energy should depend only on the total area
 $A_{{\rm tot}}=n_2 \lambda $.
The two simple coverings (with the two possible orientations) contribute
the term $2e^{-{1 \over 2}A_{{\rm tot}}}$.
There are two kinds of  double coverings depending on the
orientations of the two layers.

\leftline{{\it Equal orientations}}
 \noindent
The sum over cylindric surfaces with a cut
is given by the same formula as for the two-loop correlator, with
the only difference that here the Euler relation gives $n_0-n_1+n_2
=2$. The result is $(1-\lambda n_2+{1\over 2}\lambda^2)e^{-\lambda
A_{{\rm tot}}}
$. We have to add the term with contact interaction due to the vertex
$\omega ^{(2)}_{[1,1]}= s_{[1,1]} = (-\lambda +{1 \over 2}\lambda ^2)$
 connecting two equally oriented plaquettes:
$n_2(1-\lambda)e^{- A_{{\rm tot}}}$.

\leftline{ {\it Opposite orientations}}
\noindent
The sum over irreducible surfaces  consisting of two sheets with
opposite orientations and connected along a tree-like configuration of
links, was calculated in Appendix B of \ivn\ and
equals  $(p-2)e^{- A_{{\rm tot}}}$.
We have to add to it the term with contact interaction
due to the vertex $\omega^{(1)}_{[1,1]}=f_{[1,-1]} =-1$ connecting
 two plaquettes with opposite orientations: $-n_2 e^{- A_{{\rm tot}}}$.

Collecting all terms, we find
\eqn\enrg{\CF=2e^{-{1 \over 2} A_{{\rm tot}}}+
[-1-2A_{{\rm tot}} +
{1 \over 2}A^2_{{\rm tot}}]e^{- A_{{\rm tot}}}+ ...}

An interesting phenomenon on compact lattices was observed recently
by Douglas and Kazakov \kd . The sum over surfaces for the free energy on
the sphere is  convergent only  for sufficiently large total area
\eqn\hstytg{ A_{{\rm tot}} >\pi ^2}
At the critical point   $A_{{\rm tot}} =\pi ^2$
the entropy of the branched points overwhelms the  energy
 and the effective string tension vanishes.
The phenomenon resembles to what happens with the
compactified bosonic
string at the Hagedorn radius.

The singularity at $A=\pi^2$ can be observed already in
 the behaviour of the  multiple Wilson loops
surrounding a contour of area $A$.
{}From the explicit expression given in the Appendix
(with $\lambda$ replaced by
$A$) one finds the asymptotics
at  $n \to \infty$
\eqn\oppaaa{W_{n} (A)\approx {1 \over
\sqrt{\pi}}(n\sqrt{ A})^{-3/2}
\cos ( 2n \sqrt{A}-3\pi /4)}
The value of $A$ for which  the oscillations disappear is exactly the
critical  area  $A_{c}=\pi ^2$.
This is not a surprise since on a compact lattice with area $A$, the
free energy contains a term $ \sum _n W_n^2(A)$ (compare with eq. \ffru).

Finally, let make a  remark  concerning the case of
 a toroidal manifold $\CM$.
It was nicely demonstrated by Gross and Taylor that the free energy
of the gauge theory on the torus can be interpreted as the sum over all
connected nonfolding oriented
 surfaces wrapping the torus, with right combinatorial
factors. If we add to this sum the sum over microscopic surfaces \ffru ,
the result will be exactly the logarithm of the $\eta$-function
\eqn\trrs{\CF_{{\rm torus}}={1 \over 12}A_{{\rm tot}}
-\sum_{n=1}^{\infty}\log (1-e^{-n A_{{\rm tot}}})}

\newsec{ Strong versus weak coupling}
\subsec{The lattice string in the weak coupling phase}
The assumption that $\H_{{\rm cl}}
=\V_{{\rm cl}}=0$ is a local minimum of the free energy is justified in the
strong coupling phase of the gauge theory.
In the case of Wilson action \www\ it is known that
 the weak coupling phase  the matrix
fields will develop vacuum expectation values \jb .
The classical fields form an orbit of the gauge
group
\eqn\vev{\V_{<xy>}=\U_x \V_{{\rm cl}} \U^{-1}_y, \ \
\H_{<xy>}=\U_x \H_{{\rm cl}} \U^{-1}_y}
and are solutions of the saddle point equations
\eqn\sdpt{
\H_{{\rm cl}} = {\p S_{\lambda}(\V_{{\rm cl}}) \over \p \V_{{\rm cl}}},
 \ \ \V_{{\rm cl}} = {\p F(\H_{{\rm cl}})\over \p H_{{\rm cl}}}}
In the case of the Wilson action \www\
these equations have only trivial solution $\H_{{\rm cl}}=\V_{{\rm cl}} =0$
for $\lambda >\lambda _B$ and a nontrivial solution at $\lambda <\lambda _B$.
It is quite possible the above is true for  the general
one-plaquette action.
 In the mean field approximation the  critical coupling  $\lambda _c>
\lambda_B$ is determined by the assumption that the system chooses the
absolute maximum of the classical free energy.

Note that the saddle point equations does not fix the local
gauge transformation $\{U_x ; x\in \CL\} $.
The entropy of the gauge  degenerate mean field
\vev\ must be taken into account.
In the continuum limit it produces
 zero modes  among the oscillations around \vev\
which has to be elliminated by a gauge fixing procedure.

The diagram technique for the weak coupling phase is
obtained by expanding the effective action
$S_{\lambda}(\V) +F(\H) -{\tr \over N} \V\H$ around the classical solution
\vev
\eqn\exxxp{\H_{\l}=
 \H_{{\rm cl}} +\delta \H_{\l}; \ \
\V_{\l}= \V_{{\rm cl}} +\delta \V_{\l} }
The propagator will remain the same while new vertices will be
generated.
They will correspond to new  surface elements having one or more
free edges associated with the classical field
$ \H_{{\rm cl}}$ or $ \V_{{\rm cl}} $ while the rest of the edges are
associated with the quantum fluctuations   $\delta \V$ and $\delta \H$.
After gluing pairwise all edges labeled by the fluctuations,
the free edges will form closed loops.
Geometrically a planar diagram composed by these  elements
represents a surface with free boundaries (windows) corresponding
to these loops
 (\fig\ffafr{Microscopic structure of the world sheet
in the weak coupling phase.}).
  There are two kinds of
windows: those made out of $ \H_{{\rm cl}}$ edges
and those made of  $ \V_{{\rm cl}} $ edges along their boundary.

The gauge transformation in the classical solution \vev\ can be
absorbed in the integration measure of the fluctuations.
Therefore
the Boltzmann weights of these generalized surfaces are gauge invariant.

Let us also mention that in the weak coupling phase the potential
for the $\H$ field is given by a different expression which
is not analytically connected with the strong coupling
solution \odof \ (see Appendix A).

If the argument is proportional to the unit matrix, $\H=H\I$,
 then
the weak coupling solution gives
\eqn\wetg{F(H\I)=
2H+{\rm const}-{1\over 2} \ln H +{1\over 16 H}+{1\over 64 H^2}+...}

We have seen that in the weak coupling phase characterized by a
nontrivial classical
solution,
 windows will appear spontaneously on the world
sheet of the  string.
This situation resembles what happens in the weak coupling phase of the
vector model in which the $U(N)$
symmetry is   spontaneously broken and
which is described by  random walks  allowed to have free ends.

Whether the windows destroy or not the
world sheet in the continuum limit is a dynamical question.
Believing in the confinement, we expect that
 in $D\le4 $  dimensions
the windows are still not sufficiently large for that, and the world sheet
will  have in the continuum limit
 the structure of a dense network of thin strips
separating the windows.
 These  strips will correspond to the gluon
propagators in the standard Feynman rules.
 At the critical dimension $D=4$ the effective string
tension  of the string
 (with windows on the world sheet) should  scale with the coupling
$\lambda$  according to the
asymptotic freedom law, just as the mass in the vector model does
in $D=2$ dimensions.
 In $D>4$ dimensions the world sheet of the string is
eaten by one or several  large windows and the
Wilson loop behaves as exponential of the length of the contour.

The above qualitative picture was based on the known mean-field
solution for the Wilson action.
Whether it  is true in geleral,
  can be decided by studying the
mean field problem for a general one-plaquette action.

Such kind of  world surface  with windows
   is too   complicated as an object
 on the lattice.
However, if the string hypothesis is true,
 its critical fluctuations can be
described in terms of an embedded
 surface with continuous world sheet and measure
defined by a  local action.

\subsec{About the path integral for the continuum string}
 Our conjecture is that the
 embedded surface is defined by two fields, the coordinate $x_{\mu}(\xi),
\mu = 1,...,D$, and a field $T(\xi)$ taking its values in
 the Grassmann   manifold  $\hat G _{D,2}$ of oriented two-dimensional planes
in $\R_D$. The field $T(\xi)$ counts for the additional degrees of freedom
related to the tangent planes along  the strips.
 Put in other words, this field describes
what is left from the spin degrees of freedom
 of the virtual gluons forming the dense  planar graph.

It is convenient to parametrize the field $T(\xi)$ which has $2(D-2)$
independent components, as an antisymmetric
tensor $p_{\mu \nu}(\xi)$ satisfying a number of constraints
\eqn\consth{\eqalign{
p_{\mu \nu}p_{\mu \nu}={\rm const}, \ \ \ \ \  \
\ \ \ \ \ \ \ \ \ \ \ &  D=2,3;\cr
\ \ \ \ \ \ p_{\mu \nu}p_{\mu \nu}={\rm const}, \ \
 p_{\mu \nu}\tilde p_{\mu \nu}=0, \ \ \  & D=4\cr}
}

\eqn\consth{\eqalign{
p_{\mu \nu}p_{\mu \nu}={\rm const}, \ \ \ \ \  \
\ \ \ \ \ \ \ \ \ \ \ &  D=2,3;\cr
\ \ \ \ \ \ p_{\mu \nu}p_{\mu \nu}={\rm const}, \ \
 p_{\mu \nu}\tilde p_{\mu \nu}=0, \ \ \  & D=4\cr}
}
Following the  same logic as in the Polyakov quantization of the
 bosonic
string \ref\ppoli{A. Polyakov, \pl B (103) 207}, we can look for a local
action which is gaussian with respect to the  $x$-field  and
such that on the classical trajectories the ``intrinsic'' normal tensor
  $p_{\mu \nu}$ coincides
with the normal element $t_{\mu \nu} \sim
 \epsilon_{ab} \p _a x_{\mu} \p _b x_{\nu} $ defined by the embedding.
 Since there is no intrinsic metric, the only possible quadratic
action is given by the
  antisymmetric combination\foot{The same form of the action was recently
suggested
  by A. Polyakov in ref.
 \ref\migdal{A. Polyakov, preprint PUPT-1394, April 1993},
where the  antisymmetric field $B_{\mu \nu}(x)$ was considered as
an external field defined in the embedding space.}
  $\int d^2\xi p_{\mu \nu}(\xi)
t_{\mu \nu}(\xi)$.
The  integration measure over the Grassmann manifold will  be represented
as  a flat measure for the antisymmetric tensor field $p_{\mu \nu}(\xi)$,
with the constraints  \consth\ imposed through Lagrange multipliers
$\rho(\xi)$ and $\theta (\xi)$.
In $D=4$ dimensions the Euclidean
functional integral in the space of surfaces reads
\eqn\azza{\CD \CS =\CD x_{\mu}\CD p_{\mu \nu} \CD \rho \CD \theta
\ \ e^{-\CA}}
\eqn\act{
\CA  = \int d^2\xi[ i \epsilon_{ab}p^{\mu \nu} \p_a x_{\mu} \p_b x_{\nu}
 + {1 \over 2}\rho(p_{\mu \nu}p_{\mu \nu}+M^2)
+ i\theta  p_{\mu \nu} \tilde  p_{\mu \nu}]}
where the constant $M$ with dimension of 1/[area] is the string
 tension parameter.
This  functional integral is formally quite  similar to the
path integral for the Eguchi string \ref\egg{T. Eguchi, \prl 44 (1980) 126}.
Here the Lagrange multiplier fields restore the reparametrization invariance
which was not present in the Eguchi model.
Depending on the dimension, it is possible to add topological terms
but we are not going to discuss this point here.

Returning to the case $D=2$ we find, after integrating over the  field
$p_{\mu \nu}(\xi)=\epsilon_{\mu \nu}p(\xi)$,
the following functional  integral which strongly reminds the
path integral for one-dimensional particle:
\eqn\ppppt{\eqalign{
Z& =  \int \CD x  \CD p
 \CD \rho \exp\Big(
 - \int d^2 \xi [ i\epsilon_{ab} \epsilon_{\mu \nu} p
\p_a x_{\mu} \p_b x_{\nu}  + {1 \over 2}\rho (\xi) (p (\xi)^2 +M^2) ]\Big)
\cr
&= \int \CD x _{\mu} d\rho
\exp \Big[-{1 \over 2} \int d^2\xi   \Big( {( \p_1x_{1}\p_2x_2
-\p_1x_2 \p_{2}x_{1})^2 \over
\rho } + \rho M^2 \Big) \Bigg]\cr}
}
This analogy becomes more evident
 in the gauge $x_2=\xi_2$ in which the integral \ppppt\
is gaussian.
The classical equations
\eqn\cllas{\epsilon _{ab}\p _a p \p_b x_{\mu}=0, \ \ \
p=i \epsilon_{ab} \epsilon_{\mu \nu}
\p_a x_{\mu} \p_b x_{\nu}  / \rho = \pm iM}
allow discontinuities of the $p$-field which correspond to embeddings
having branch points and
 folds.
Whether to take into account or not the configurations with folds
depends on the regularization of  the path integral \ppppt .

 The loop correlators defined with such a measure are Stokes functionals
in the sense of ref.\mm ,
i.e., they have
 a nonsingular area derivative, which  property is shared by the
(nonrenormalized) Wilson loops in the gauge theory.

\newsec{Discussion}

We formulated  a model of random surfaces (lattice strings)
satisfying the backtracking condition \unn\
and the lattice analogue of the loop equation \mme .
This model describes the strong coupling phase of the $U(N)$ gauge theory
in the limit $N\to\infty$ as well as the $1/N$ corrections.

In the simplest case of two dimensions
 this lattice  string model is equivalent to
the gauge theory up to the continuum limit.
 Nevertheless  this
  is still not a continuum string
theory with a local action on the world sheet.
The most important step towards the continuum theory, the translation of  the
random surface
{\it Ansatz} from the language of cellular complexes to the language of
differential forms, is still to be done.
The main problem to solve
 is to reformulate the contributions of the puncture operators
in terms of  local fields  defined on the
world sheet.
In other words,  we have to look for
 a continuum version of the puncture operators
\aaapa\ and \vvvpv \ defined on the 1-cells and 2-cells of the lattice
using the language of differential forms.

The weights \omom\ of the branch points consist of a
dimensionless ``topological'' term
and a term proportional to the infinitesimal area $\lambda$.
(The higher powers in $\lambda$ can be neglected in the continuum
limit.)
The term proportional to $\lambda$ is easy to interpret in the
continuum limit: it corresponds to a point-like
singularity of the intrinsic curvature
weighed by a factor $(-1)$.
The factor $\lambda$ comes from the measure over surfaces,
as we have noticed in \ivn .

The real problem consists in the interpretation of the
dimensionless weights. They seem to be associated with
puncture operators
which are pure derivatives.
These are the only singularities
 in the ``topological'' limit $\lambda =0$
of the model. In this limit  the vector potential  is restricted
to be a pure gauge,  $A_{\mu}(x)= U^{-1}(x)\p_{\mu}U(x)$.
In this ``topological'' string theory
the sum over surfaces bounded by a loop
 is always 1.

Presumably there can exist only a finite number
of local excitations with the lowest dimension.
We have to study the classes of universality of local operators on
the world sheet to extract what is left from the puncture
operators in the continuum imit.

\bigbreak\bigskip\centerline{{\bf Acknowledgements}}\nobreak
The author thanks M. Douglas, D. Gross, V. Kazakov, A. Migdal,
A. Polyakov and A. Tseytlin
for stimulating discussions, and
especially  J.-B. Zuber for his critical remarks.

\appendix{A}{ Expansion of the potentials in terms of moments}

\subsec{The potential for the $\H$-field}
The function $F$ can be determined using the Ward identity
\eqn\uniiy{ [\tr (\p /\p \H^{\dag} \p / \p \H)-N]e^{N^2F(\H)}=0}
 which is the matrix analogue of \wwwar .
In the  large-$N$ limit this function  was calculated explicitly
by Brezin and Gross \ref\brgr{E. Br\'ezin
and D. Gross, \pl 97B (1980) 120}.
 in terms of the eigenvalues of the
Hermitean matrix $\H^{\dag}\H$.
The result is given by  two different  analytic
expressions  depending on the strength of the
external field $H$. These correspond to two different
saddle points of the integral \onlin .
In the  ``strong coupling''
 regime of small field $\H$
  characterized by the condition
$\tr (\H^{\dag}\H)^{-1/2}>2N$ the solution is
\eqn\doiu{F(\H)={2 \over N}\sum_{i=1}^N \sqrt{\lambda_i+c}-
{1 \over 2N^2}\sum_{i,j} \log (\sqrt{\lambda_i+c}+\sqrt{\lambda_j +c})
-c-{3 \over 4}}
where $\lambda_i, i=1,2,...,N$ are the eigenvalues of the Hermitean
 matrix
$\H^{\dag}\H$
and the parameter $c$ is determined by
\eqn\cccc{ {\tr \over N} {1 \over \sqrt{\H^{\dag}\H +c\I} }=
{1\over N}\sum_{i}{1\over\sqrt{\lambda_i+c}}=2}
In the ``weak coupling'' regime defined by the condition
 $\tr (\H^{\dag}\H)^{-1/2}<2N$
the solution is given by the same formula \doiu\ with $c=0$.

For sufficiently small field $\H$ the potential $F(\H)$
 can be expanded as a series in the
momenta
\eqn\mmmi{\a_n={ \tr \over N} (\H^{\dag}\H)^n,}
\eqn\odof{ F[\a]=\sum_{n=1}^{\infty}
\sum_{k_1,...,k_n \ge 1}
f_{[k_1,...,k_n]}{\a_{k_1}
...\a_{k_n} \over n!}}
The coefficients of the series were found in the large $N$ limit
by O'Brien and Zuber \ref\objb{K.H. O'Brien and J.B. Zuber, \pl 144 B (1984)
407}
as a solution of
a system of algebraic relations
suggested originally  by Kazakov in \kaza .
These relations are equivalent to  eq. \uniiy\ written in terms of loop
variables
\eqn\uniyy{
\Big( \sum_{n\ge 1}n\a_{n-1}\p_n+
\sum_{k,n \ge 1}[ (n+k+1)\a_n\a_k\p_{k+n+1}+
 {nk\over N^2} \a_{n+k-1}\p_n \p_k ] -1
\Big)\ e^{N^2F[\a]}=0}
where we denoted $\a_0=1, \p_n=\p / \p \a_n$.
Eq. \uniyy\
is sufficient to determine   the   $1/N$ expansion of the link-vertices
$f_{[k_1,...,k_n]}$.
In the large $N$ limit the identity \uniyy\ is equivalent to a
 system of recurrence relations \obrz
\eqn\oopol{\eqalign{
f_{[1]}&=1\cr
f_{[k_1,...,k_n]}+\sum _{j=2}^{n} k_j f_{[k_1+k_j,k_2,...,\bar k_j,
...,k_n]}
+\sum_{k=1}^{k_1 -1} \sum _{\alpha \in P(L)} f_{[k_1-k, L\backslash \alpha]}
f_{[k,\alpha]}&=0\cr}}
where $L={k_2,...,k_n}$ and $P(L)$ is the set of all subsets of $L$,
including the emply set, and the bar means omitting the argument below it.
The equations for the  weights $f_{[k]}$ of the disks \kaza
\eqn\plopp{f_{[1]}=1; \ \
f_{[n]}+\sum_{k=1}^{n-1} f_{[k]}f_{[n-k]}=0, \ n=1,2,...}
are solved by the Catalan numbers
\eqn\ihy{f_{[n]}=(-)^{(n-1)}{(2n-2)! \over n! (n-1)!}}

 \subsec{The potential for the $\V$ field}

The potential $S_{\lambda}(V)$ is defined by the analytic continuation
of the r.h.s. of \chpo\ from the hyperplane $\V=\U,\ \  \U^{\dag}\U=\I$ to
the space of all complex matrices $\V$.
For this purpose one has first to express the characters
\chhr\ in terms of the moments of the
matrices $\V$ and $\V^{\dag}$ by replacing $u_i^n \to v_i^n, u^
{-n}_i\to
\bar v^n_i$.
After that the
r.h.s. of \chpo\ can be written as a series in  the moments
\eqn\mimmi{ \beta _n=
{ \tr \over N} \V^n , \  \beta _{-n}={ \tr \over N}
  \V^{\dag n}
;\ \ \ \  \ n=1,2,... }
\eqn\asda{N^2 S_{\lambda}[\beta]=N^2 \sum_{n=1}^{\infty}
 \sum_{k_1,...,k_n \ne 0}
s_{[k_1...k_n]} {\beta _{k_1}...\beta _{k_n}
\over n!}}
The coefficients in the expansion  \asda\ of the heat kernel action
 can be found order by order
by taking the logarithm of this series
\eqn\chppo{\eqalign{
e^{N^2 S_{\lambda}(\V)}&=\sum_{\vec n}\bar \chi_{\vec n}(\I)
 \chi_{\vec n}(\V)
e^{-\lambda (2N)^{-1}C_{2}(\vec n)}\cr
&=1+N
(\tr \V+\tr \V^{\dag}) e^{-\lambda /2}+(N^2-1)
(\tr \V
\tr \V^{\dag} -1)e^{-\lambda} +...\cr
}
}
\eqn\ccsc{\eqalign{
S_{\lambda}(\V)&= e^{-\lambda /2} (\v_1+\v_{-1})
-e^{-\lambda}(\v_1\v_{-1}+N^2) \cr
&+
{1 \over 2}    (\v_2+\v_{-2})(\cosh {\lambda \over N}
-N\sinh {\lambda \over N})e^{-\lambda}+
{1 \over 2!}(\v_1^2+\v_{-1}^2)(-N\sinh {\lambda \over N}
\cr &+ 2 (N\sinh {\lambda \over 2 N})^2)e^{-\lambda}+...
\cr}
}
which gives, in the limit $N\to\infty$,
\eqn\cooe{\eqalign{
s_{[1]}&=e^{-\lambda /2},\cr
 s_{[2]}&=
(\cosh {\lambda \over N}
-N\sinh {\lambda \over N})e^{-\lambda} =
(1-\lambda)e^{-\lambda}+\CO({1 \over N^2})\cr
s_{1,-1}&= -e^{-\lambda}\cr
s_{[1,1]}&= (-N\sinh {\lambda \over N}
+ 2 (\sinh {\lambda \over 2 N})^2)e^{-\lambda}
=(-\lambda +{1 \over 2}\lambda ^2)e^{-\lambda}+\CO({1 \over N^2})\cr
}
}

In the continuum limit $\lambda \to 0$ the expansion $\asda$
takes asymptotically the form
\foot{This formula resembles eq. (22) in the paper by M. Douglas \dg .
We tried to establish an exact correspondence between his
formalism and ours, but haven't succeed in this}
\eqn\ehaa{S_{\lambda =0}(\beta)=\sum _{n=1}^{\infty}{1 \over n}
 (\beta_n +\beta_{-n} - \beta_n\beta_{-n} ) = \prod _{i,j-1}^N
{1-v_i\bar v_j \over (1-v_i)^N(1-\bar v_j)^N}
}
which tends to the  $\delta$-function $\delta (\V,\I)$
 when $\V$ approaches the hypersurface $\V^{\dag}\V=\I$.

The coefficients of the expansion \asda\ satisfy a system of loop equations
equivalent to the heat kernel equation on the $U(N)$
group manifold   \hhhhk\ projected into the space of
nonrestricted complex matrices
\eqn\hke{\Big[ 2N{\p \over \p \lambda} + \tr \Big(
\V\p /\p \V - \V^{\dag}\p /\p \V^{\dag}
\Big)^2 \Big] e^{N^2 S_{\lambda}(\V)} \ = \ 0}
Namely,
\eqn\hkee{\Big( 2 \p / \p \lambda +
\sum_{n\ne 0} |n|\v_{n}\p_n+
\sum_{k,n \ne 0}[\theta(nk) |n+k|\v_n\v_k\p_{n+k}+
{ nk\over N^2} \v_{n+k}\p_n\p_k]
\Big)\ e^{N^2S_{\lambda}[\v]}=0}
where we used the
shorthand notation $\p_n=\p/\p \v_n$.

This identity allows in principle to determine the coefficients
$s_{[k_1,...,k_n]}$
 in the expansion of $S$ in the loop variables,
but it is not triangular and does not produce simple recursion relations.
Such recursion relations will be written below for the set of
all possible loop correlators
 \eqn\connesi{W_{k_1,...,k_n}=
N^{n-2} \langle \tr U^{k_1} ... \tr U^{k_n} \rangle _{
{\rm connected}}}

 The coefficients $s_{[k_1,...,k_n]}$ are related to the elementary
Wilson loop correlators \connesi\ by
\eqn\pish{s_{[k_1,...,k_n]} = W_{k_1,...,k_n} , \ \ \ \ n \ne 2}
and, for $n=2$
\eqn\hhy{s_{[k_1,k_2]}= W_{k_1,k_2}- \delta _{k_1 , -k_2}}
In particular, $s_n =W_n$, as it has been noticed in \ivn .

\subsec{The one-plaquette Wilson loops}
Let us denote by
\eqn\jjji{\eqalign{
 W_{\lambda}  [x]
& ={1\over N^2}  \log  \int \CD U \exp [N^2  S_{\lambda}(U) +N\sum_{n\ne 0}
y_n \tr U^n ]\cr
&= \sum _n N^{2-n}{1 \over n!} \sum_{k_1,...,k_n \ne 0}
W_{k_1,...,k_n}x_{k_1}...x_{k_n}\cr}
}
the generating function for the connected  Wilson loop correlators
\connesi .

Applying the identity \hke\ and integrating by parts in \jjji , we find
the   loop equation
\eqn\looki{\eqalign{
\Big[
2{\p \over \p \lambda}&+\sum_{n\ne 0} |n| x_n \p_n+
 \sum_{n,k \ne 0}  \Big(
n \ x_n \ k \  x_k \ \p_{n+k} \cr
 &+{1 \over N^2}\theta (kn)
 (|n|+|k|) x_{n+k} \p_n\     \p_k +
\Big)\Big] e^{N^2W[x]}
 =0 \cr}}
which is in a sense dual to \hkee .
  This equation is a particular case of the Migdal-Makeenko equations
derived for a general Wilson loop in \kk .

In the limit $N\to \infty$ the loop equation \looki\
yields  a system of recursive equations for the connected correlators
$W_{k_1...k_n}$
\eqn\ftap{\eqalign{
(2{\p \over \p \lambda} +\sum_{i=1}^{n} k_i)W_{k_1,k_2,...,k_n}&+
{1\over N^2}
\sum _{i\ne j}^{n} k_i k_{j} W_{k_i+k_j, k_2,...,\bar k_j, ...,
k_n}\cr &+\sum_{i=1}^{n}\sum _{k=1}^{k_i-1} \sum_{\alpha \in P(L)}
|k_i| W_{k_i-k, L\backslash \alpha}W_{k, \alpha}
=0\cr}}
In particular, for the simple Wilson loops
$W_{n} = (1/N)\langle \tr U \rangle$, eq. \ftap\ reads
\eqn\iioip{
2{\p \over \p \lambda}W_n+ n \sum _{k=1}^{n}W_{k}W_{n-k}=0}
This equation has been studied in
 \ros . The solution of \iioip\ is
\eqn\polop{
W_{n}=e^{-n\lambda /2}\sum_{m=0}^{n-1} {(-n\lambda)^m
\over (m+1)!}=e^{-n\lambda /2}(1-{n(n-1)\over 2}\lambda + ...)}

\listrefs
\listfigs
\bye
We shall see that the  vacuum energy $\log Z$ can be written as a sum
over connected closed surfaces  embedded in $\CL$. The elementary cells of
these surfaces represent multiple coverings of the elementary cells of
the lattice $\CL$, i.e., they  can contain branch points.
Similarly, there are branch points associated with the links of $\CL$.
The model is slightly more complicated than the Weingarten model where
branch points can appear only at the sites of the lattice, but this is
the price payed for not having  tachyons.

The  Boltzmann weights
are products of local factors  associated with  the links and cells
(plaquettes)
of the lattice surface.
Even in the planar limit this is
 not yet a model of noninteracting random surfaces.
Indeed, the generic vacuum surface looks as a cluster of spheres
glued together by means of nonplanar surface elements
(\fig\figiv{A generic surface contributing to the
vacuum energy in the planar limit}).
Looking at a time slice of \figiv\ we see that the nonplanar surface elements
describe  the interaction of the QCD string with virtual vacuum surfaces.
This tadpole interaction means that we did not expand about the true
string vacuum, and   can be removed by  a redefinition of the fields.
(That is, the  field $V$ in eq. \oio\ has to be given a nonzero
classical value.)

If we expand  the interaction potentials
in the loop variables

and

then the coefficients in the expansion
will be the local weight factors corresponding the the elements of the
random surface. They
represent infinite series in $1/N^2$
\eqn\gtf{\eqalign{
f_{[k_1,...,k_n]} &=\sum_{m=0}^{\infty}
N^{-2m}f^{(m)}_{[k_{1},...,k_n]}
\cr
g_{[k_1,...,k_n]} &=\sum_{m=0}^{\infty}
N^{-2m}g^{(m)}_{[k_{1},...,k_n]}
\cr}}
and can be determined from simple string equations which will be derived in the
next section.

The integral over
the matrix fields
yields the sum over all coverings of the lattice $\CL$  weighed with
products of $z^*$ and $t^*$ fields.

 surface elements  with the topology of a disk.
In fact the Boltzmann weights  in the truncated  sum over random surfaces
coincide with the original ones. This follows from the fact that
the contributions of the surfaces renormalizing the type-1 and type-2 vertices
vanishes due to the unitarity of the original theory.
In ref.\ivn\ they are called plaquette- and link-vertices.

A branched covering of the lattice $\CL$ is
defined as a discretization of a branched covering of the continuous
manifold by a two-dimensional surface.
The general position of a branch point is on the 2-cells $c$, but it can
occur also along the links $\l$ or at the sites $s$ of the lattice.
After being  cut along the links of $\CL$, the surface
  decomposes into a set of elementary oriented  surfaces representing
branched coverings of its 2-cells.
To make connection with the traditional language used in lattice gauge
theories, we refer to
a connected elementary surface representing $n$-covering of a cell $c\in \CL$
as a {\it multiplaquette } of order $n$, or simply $n-plaquette$
 (\fig\figi{Multiplaquettes (branched
 coverings of a 2-cell) of orders 1,2,3.}). An

 where $\p c$
is the boundary of the cell $c$.
It has
 a  branched point of order $n$
and    area
 $n\lambda$.

The multiplaquettes will serve as
building blocks  for  the world sheet of the lattice string.

A branched covering of the complex $\CL$ can contain also branch points
trapped in  its 0-cells (sites) $s$ and 1-cells (links) $\l$.
A branched point associated  with a link $\l$ means a cyclic contraction of
 the half-edges of
the multiplaquettes located at this link.
Such a contraction  can be visualized (V. Kazakov, \kaza )
 as a piece of a surface with zero area and a boundary
 $(\l \l^{-1})^n$ (\fig\figii{Kazakov saddles (branched coverings
of a 1-cell) of orders 1,2,3}).

For each  closed contour in $\CL$, the  Wilson loop average $W(C)$
 is given by the sum over all  minimal  branched
surfaces \foot{The folds are not forbidden but their contributions
cancel} of $\CL$ bounded by $C$.
The weight of each surface is equal to exp [ - area] times a product of
local factors,  $g_n$ and $f_n$,
 associated with its branch points.  Note that the
 weight  of a branched point depends on whether is is located at
a 0-, 1-, or 2-cell.

A branched point on the world sheet describes a local exchange interaction
between several strings.  The fact that its amplitude depends on
the dimension of the cell where it occurs urges to be given meaning in the
continuum.

This simplest form of the random surface {\it Ansatz} applies only to the
one-loop average in the limit $N=\infty$. The string path integral
representation of the multi-loop amplitudes or the free energy
involves surface elements with more complicated topology.
These surface elements are again located at the links and cells of the lattice
and have the topology of a sphere with a number of boundaries and
handles.
They describe all possible string interactions containing virtual
closed string states.  In this paper we
 generalize the random surface {\it Ansatz} by
including these elements in the sum over surfaces and calculate
their Boltzmann weights.
Then we demonstrate how the {\it Ansatz} works on some simplest examples
and reproduce the results obtained by the character expansion method
or the loop equations.

 It seems  likely that, just as in the case of the
 one-dimensional particle, the configurations with folds only renormalize
the string tension $M^2$ and  the classical solution giving the
nonoriented area yields the exact result.
In fact, this is what one expects in the usual Weingarten model:
the sum over surfaces is dominated by surfaces of minimal area spanning
the loop and developing a number of ``baby universes''. The critical
singularity appears due to the explosion of the baby universes and
not due to some long range fluctuations of the world sheet.
Therefore, the entropy of surfaces can only renormalize the string
tension and the loop average is just the exponent of the minimal area.

will be described by the embedding $\xi \to x_{\mu}(\xi)$ as well us by

Before trying to make this step we need to know what  the world sheet of
the continuum QCD string represents by itself.
As it was first realized  by Polyakov \ref\ppol{\pl 103B (1981) 207},
the intrinsic geometry of the quantum surface is related to
the embedding only at dynamical level. In the case of the
bosonic string
the  intrinsic metric  is kinematically
independent from  the induced one.

Here we propose a possible measure in the space of surfaces which
seems to be the one relevant for the QCD string.
The r\^ole of the intrinsic metric of the world sheet  is played here
by a momentum  field  defining an ``intrinsic'' tangent plane at each point of
the world sheet,
A very similar idea

Returning to the case $D=2$ we find, after integrating over the  field
$p_{\mu \nu}(\xi)=\epsilon_{\mu \nu}p(\xi)$,
the following functional  integral which strongly reminds the
path integral for one-dimensional particle:
\eqn\ppppt{\eqalign{
Z& =  \int \CD x  \CD p
 \CD \rho \exp\Big(
 - \int d^2 \xi [ i\epsilon_{ab} \epsilon_{\mu \nu} p
\p_a x_{\mu} \p_b x_{\nu}  + {1 \over 2}\rho (\xi) (p (\xi)^2 +M^2) ]\Big)
\cr
&= \int \CD x _{\mu} d\rho
\exp \Big[-{1 \over 2} \int d^2\xi   \Big( {( \p_1x_{1}\p_2x_2
-\p_1x_2 \p_{2}x_{1})^2 \over
\rho } + \rho M^2 \Big) \Bigg]\cr}
}
(This becomes evident in the gauge $x_2=\xi_2$ in which the integral \ppppt\
is gaussian).
The classical equations
\eqn\cllas{\epsilon _{ab}\p _a p \p_b x_{\mu}=0,
p=i \epsilon_{ab} \epsilon_{\mu \nu}
\p_a x_{\mu} \p_b x_{\nu}  / \rho = \pm iM}
allow discontinuities of the $p$-field which correspond to embeddings
having branch points and
 folds.
Whether to take into account or not the configurations with folds
depends on the regularization of  the path integral \ppppt .

**********************************************************************

\subsec{Feynman rules in the large $N$ limit}

The Feynman rules are obtained as follows.

Inverting the bilinear part of the action we find the propagators
\eqn\connt{\langle  (H_{\l})^i_j (V^{\dag }_{\l})^k_l \rangle =
\langle  (H^{\dag}_{\l})^i_j (V_{\l})^k_l \rangle
={1 \over N}\delta^i_l \delta^{k}_j ; \
\langle   (V^{\dag}_{\l})^i_j (V_{\l})^k_l \rangle
       ={\ta_1 \over N}\delta^i_l \delta^{k}_j
}
 We associate with the vertex
 $\tilde \beta _{n}(p) \tr \V^{n}_p$
 a little surface with the topology of a punctured
disk having as a  boundary the loop  $(\p p)^n$
 going $n$ times around the boundary $\p p$
of the elementary plaquette
$p$. The disk connects the operator $ \tr \V^{n}_p$
associated with the boundary with the operator
 $\tilde \beta _{n} $ associated with the puncture.
   This surface covers $n$ times the plaquette $p$ and we
will refer to it as $n$-plaquette.
The first three $n$-plaquettes  ($n=1,2,3$) are shown in
 fig. \fig\fmpl{$n$-plaquettes, $n$= 1,2,3}  where for
 eye's convenience we have displaced the edges of the boundary.
We assume that the  Gaussian curvature $2\pi (n-1)$
is concentrated at the puncture.

The $n$-plaquettes will serve as building blocks for constructing the
world sheet of the lattice string.
The  Boltzmann weight of an $n$-plaquette in $\tilde \beta_n$.
The propagator $\V-\V$ has simple geometrical meaning: it glues two edges
of multiplaquettes having opposite orientations.

The absence of a $H-H$ propagator means that the $\H$ field plays only  the
role of glue for identifying the edges of the multiplaquettes.
Besides the simple contraction $\V-\V$ there are cyclic contractions
of $n$ pairs of oppositely oriented edges, $n=2,3,...$,
which are represented by
the vertices  $\ta_n(\l)  \tr (\H^{\dag}_{\l}\H_{\l})^n$.
As before, we associate with the operator $\ta_n(\l)$
a puncture  where a curvature $2\pi (n-1)$ is concentrated.
The edges involved in such cyclic contraction are glued half-by-half.
This geometrical image appeared first in the paper by Kazakov \kaza .
The simple contraction and the first two  nontrivial  ones are shown in fig.
\fig\fcci{Cyclic contractions (Kazakov saddles) involving
 2$n$ edges, $n$=1,2,3}

The vacuum Feynman diagrams  are in one-to-one correspondence
with  surfaces composed of $n$-plaquettes
glued together along their half-edges.

An important feature of  these surfaces is that they can
have local singularities of the
curvature not only at the sites, as it is the case in the original
Weingarten model, but also at the links and plaquettes  of the surface.
These singularities can be interpreted as processes of splitting
and joining of lattice strings.
The weight of a closed surface is a product of the weights of its
elements times a power of $N$ which, with our normalizations, is
equal to the Euler characteristics of the surface.

The free energy of the $U(\infty)$ gauge theory is given by the sum over
all closed connected surfaces with the topology of a sphere.

The connection with the traditional Feynman-like diagrams is by duality.
Sometimes it is mire convenient to use the traditional
diagrammatical notations. Then a surface bounded by a loop
$\Gamma = [\l_1\l_2...\l_n]$ will be represented
by a planar Feynman graph with $ n$ external legs dual to the links
$\l_1,...,\l_n$.
The cyclic contraction of $2n$ is represented by a vertex with $2n$ lines
as is shown in fig. 4.

\subsec{Evaluation of the weights of the  cyclic contractions}

Consider  the Wilson loop average for the
 contour $\Gamma=(\l_1\l_2...)
$ with coinciding endpoints.

\subsec{String representation of the Wilson loop average.
Resum\'e}

Let us summarize what we have achieved by now.
The Wilson loop average $W(\Gamma)$
 in the $U(\infty)$ gauge theory defined on
a $D$-dimensional lattice is equal to the sum of all planar
 irreducible surfaces having as a boundary the loop $\Gamma$.
These surfaces are allowed to have branch points
  of all orders
located at the sites, links and plaquettes of the space-time lattice.
Introducing unified notations $\omega_{n}^{(k)}$
 for the weights of the  branch points associated with
the $k$-cells of the space-time lattice (sites are 0-cells, links ate
1-cells, and plaquettes are 2-cells)
where
\eqn\omom{\eqalign{
\omega^{(0)}_ n
&= 1\cr
\omega^{(1)}_n &= f_{[n]}=(-)^{(n-1)}{(2n-2)! \over n! (n-1)!}\cr
\omega^{(2)}_n &= {\tilde \beta_n \over (\tilde \beta_1)^n}
}}
we can write the sum over surfaces as
\eqn\willsl{W(\C)=
\sum_{ \p \CS =\C} \Omega (\CS) e^{-M_0 {\rm Area (\CS)}}}
where
the factor $\Omega (\CS)$ is a product of the weights of all branched points
of the world sheet and $M_0= - \ln \tilde \beta_1$.

\subsec{The contribution of the surfaces with folds is zero}
The weights of the branch points
 provide a mechanism of suppressing the backtracking
motion   of the strings or, which is the same, world surfaces having folds.
The  surfaces with folds have both positive and negative weights and
their total contribution to the string path integral is zero.
We have checked that in many particular cases but the general proof is missing.

It is perhaps instructive to give one example. Consider the
surface in fig. 7.
It  covers three times the interiour of the nonselfintersecting
loop $C$ and once - the rest of the lattice.
There are two specific points on the loop $C$ at which the curvature
has a conical singularity; they can be thought of as the points
where the fold is created and annihilated.
Each of these points can occur either at a site or at a link (in the
last case it is associated with a weight factor $\omega^{(1)}_2=f_{[2]}=-1$).

Let us evaluate the total contribution of all irreducible  surfaces
 distinguished by the positions of the two singular points.
Denoting by $n_0$ and $n_1 (=n_0)$
 the number of sites and links along the loop $C$,
and remembering that the a singular point located at a link has to be taken
with a weight $
f_{[2]}=-1$, we find that the contribution of these surfaces
is proportional to
\eqn\ffafa{[{n_0(n_0-1) \over 2}+f_{[2]}n_0n_1+f_{[2]}^2{n_1(n_1-1)\over 2}]
-[n_1 +2 f_{[2]}n_1] =0}
The second term on the left hand side contains the contribution of the
reducible surfaces which had to be subtracted.
A reducible surface arises when the two points are located at the extremities
of the same link ($n_1$ configurations) or when one  of the points is
a   branch point and the other is located at one of the extremities of the
same link ($2n_1$ configurations).

\newsec{ The trivial  $D=2$ gauge theory as a nontrivial
model of random surfaces}

Let us consider the formal path integral of the $U(N)$ gauge theory on the
oriented two-dimensional manifold  $\CM$
\eqn\i{\CZ=\int \prod _{x\in \CM}dA_{\mu}(x)\exp[-{1 \over 2 }N
\int  \tr  F^2 d^2x]}
where the gauge field $A_\mu (x)$ belongs to the Lie-algebra of $U(N)$
and $F= \p _{1} A_{2} - \p _{2} A_{1} + [A_{1},A_{2}]$
 is the field strength.
Choose a  lattice $\CL$ covering the manifold   $\CM$ and having the structure
of a two-dimensional simplicial complex consisting of
plaquettes  $p$, links $\l$ and sites $s$.
Assume for simplicity that all cells have the same area $\lambda $.
After performing the integration over the gauge field inside the cells $c$,
the integrand will depend only on the  link variables
  \eqn\kkak{U_{\l}=\exp[i \int _{x_i}^{x_j}dx_{\mu}A_{\mu}(x)]}
associated with the links $\l =<s_i s_j>$ of the lattice and belonging to
the fundamental representation of the group $U(N)$.
The result of the integration in each cell $c$ depends only on the holonomy
factor associated with its boundary $\p c$
\eqn\plaq{U_c=
\exp[i\oint _{\p c}dx_{\mu}A_{\mu}(x)]
=\prod_{\l \in \p c} U_{\l} }
and equals  the heat kernel on the  $U(N)$ group manifold\foot{This can be
proved easily in the axial gauge  $A_2=0$. Let the  interior of the cell $c$
be the set if the points with coordinates $(x_1,x_2)$  such that
$x_1^d\le x_1 \le x_1^u, x_2^l(x_1)\le x_2\le x_2^r(x_1)$.
Then we can impose an additional gauge condition $A_1(x_1,x_2^l(x_1))=0$.
The
  functional integral over $A_1$ is gaussian and is given by the
exponentional  of the classical action $-1/2\int_c d^2x (\p_2A_1)^2$
where $A_1= (x_2-x_2^l(x_1))U^{-1}(x_1)\p_1U(x_1)$ and $U^{-1}(x^u_1)
U(x^d_1)=U_c$. The functional
integral over the residual
 degree of freedom $U(x_1)$
is equal to the partition function of a one-dimensional $U(N)$ chiral field
defined on the interval $0\le \tau = \int ^{x_1} ( x_2^r(x_1)-x^l_2(x_1))
 dx_1 \le \lambda$ with boundary condition $U^{-1}(\tau)U(0)=U_c$}
\eqn\deis{e^{ S_{\lambda} (U)}= \langle I |e^{{\lambda \over 2N}
 \hat  C_2 \p /\p U)^2 }|U \rangle
= \sum_{R}\bar  \chi_{R}(I) \chi_{R}(U)e^{-{\lambda A \over 2N}C_{2}(R)}
}
The sum in \deis\ goes over all irreducible representations $R$ of $U(N)$
and $C_{2}(R)$ is the Fourier image of the Laplacian $\hat C_2
= \tr (U^{-1}\p /\p U)^2$  (the quadratic Casimir
operator).
After that the  partition
function \i\   reduces to
\eqn\discr{
Z(\CL)= \int _{\l \in \CL} [dU_{\l}]
\exp[\sum_{c \in \CL}S_{\lambda}(U_ c)]}
where $[d U]$ is the invariant measure on the group $U(N)$.

The case $D=2$ was considered recently in details in \ref\az{I. Kostov,
Saclay preprint T93/050, 1993; submitted to \np B}.
For the heat kernel action the string tension $M_0$
and the  weights $\omega_n^{(2)}$ are
given by
\eqn\wowop{
M_{0}= {\lambda \over 2}; \ \
\omega_n^{(2)}= \sum _{m=1}^{n-1}\pmatrix{n\cr
m+1\cr}{n^{m-1} \over m!}(-\lambda)^m
=(1-{n(n-1)\over 2}\lambda + ...)}

\subsec{String representation for the Wilson loops on the infinite plane.}

Now we are able to formulate the string {\it Ansatz} for the
Wilson loop average  as the sum over all
 irreducible surfaces without folds, spanning the loop $C$
and having branch points of all orders
\eqn\willsl{W(C)=
\sum_{ \p \CS =C} \Omega (\CS) e^{-{1 \over 2} {\rm Area (\CS)}}}
The factor $\Omega (\CS)$ is a product of the weights of all branched points
of the world sheet.
A branch point of order $n$ ($n=1,2,3,...$)
associated
 with a $k$-cell ($k=0,1,2$
  is weighed by a factor
$\omega_{n}^{(k)}$ where
\eqn\omom{\eqalign{
\omega^{(0)}_ n
&= 1\cr
\omega^{(1)}_n &= f_{[n]}=(-)^{(n-1)}{(2n-2)! \over n! (n-1)!}\cr
\omega^{(2)}_n &= s_{[n]}e^{n\lambda/2} =\sum_{m=0}^{n-1}
\pmatrix{n\cr
m+1\cr}{n^{m-1} \over m!}(-\lambda)^m
=(1-{n(n-1)\over 2}\lambda + ...)\cr}
}
\subsec{Wilson loops on the infinite plane. Examples}

Let us illustrate how the
string {\it Ansatz} \willsl \ works by  some simplest
   examples of planar loops (\fig\ffig{Examples of Wilson loops:
a) Circle, b) Flower, c) Pochhammer contour}) which have been calculated
previously using the Migdal-Makeenko loop equations \kk .

\leftline{$a)$ {\it Circle}}
Assume that the loop encloses $n_2$  2-cells of area $\lambda$.
The sum over surfaces contains a single
 nonfolding surface spanning
the loop and
\eqn\siml{W(C)= e^{-{1 \over 2}n_c\lambda}=e^{-{1 \over 2}A}}
where $A$ is the area enclosed by the loop.
This  is the famous area law for the Wilson loops in ${\rm QCD}_2$.

\leftline{$b)$ {\it Flower}}

The nonfolding surfaces spanning the loop will cover the three petals
of the flower (denoted by 1 in \ffig ) once, and the head (denoted by 2)
- twice. Therefore the total area will be always $A= A_1+2A_2$,
where $A_i$ is the area enclosed by the domain $i \ (i=1,2)$.
Each of these surfaces will have a branch point located at some site,
link or cell of the overlapping area 2 \fig\faaa{A branched covering of
the flower contour}.
Let us denote by $n_{0}, n_{1 }, n_{2 }$ the numbers
of sites, links, cells belonging to the domain 2.
Then the sum over the $\Omega$-factors due to the branch points
 reads
\eqn\omfac{\sum \Omega =
\sum_{k=0,1,2}n_{k}
\omega_2^{(k)} = (n_0-n_1+(1-\lambda)n_2)=1-n_2\lambda}
(Here we used the Euler relation $n_0-n_1+n_2 =1$.)
Therefore
\eqn\womg{W(C)=(1-A_2) e^{-{1 \over 2}(A_1+2A_2)}}

\leftline{$c)$ {\it  Pochhammer contour}}

This time the surfaces spanning $C$  may have different area.
The three possible cases are shown in \fig\fffirg{Three ways of
spanning the Pochhammer contour}.
There is one surface with area $A_1+A_{\bar 2}$, one surface with area
$A_1+A_2$ and a whole class of surfaces with area $A_1+A_2+A_{\bar 2}$
which are distinguished by the position of a branch point of order $2$.
The two overlapping sheets are with opposite orientations and the
branch point can occur only along the boundary of the overlapping
region. Let $n_0$ and $n_1$ be the numbers of sites and links belonging to
this boundary. By the Euler relation, $n_0-n_1 =1-2=-1$ (there are two cycles).
Therefore the sum over the positions of the branch point contributes a
factor $\sum _{k=0,1} n_k \omega_2^{(k)} =n_0-n_1=-1$
and  the Wilson average reads
  \eqn\ovoma{W(C)= e^{-{1\over 2}(A_1+A_2)}+e^{-{1\over 2}(A_1+A_{\bar 2})}
-e^{-{1\over 2}(A_1+A_2+A_{\bar 2})}}

\smallskip

Finally, consider  the sum over vacuum surfaces in the large $N$ limit.
It follows from \clfe\ that the  vacuum energy is equal to the  sum over
surfaces  that, after being cut along all copies of a  given link, split
only into irreducible pieces.
On the infinite plane, the only surfaces
satisfying these  conditions are made by identifying the boundaries of two
multiplaquettes and the free energy  is given by
\eqn\ffru{\CF= (\# {\rm cells}) \sum _{n=1}^{\infty} {s_n^2\over n}}
In the continuum limit $\lambda \to 0, A_{{\rm tot}}
=\lambda \# {\rm plaquettes }$ = const,
we find, up to an infinite constant,
\eqn\okok{\CF = { A_{{\rm tot}}\over \lambda}\sum _{n=1}^{\infty}
{1 \over n}(1-n^2\lambda + ...) = {1 \over 12} A_{{\rm tot}}}

\subsec{Interactions}

The simplest example where the interactions appear is the two-loop
correlator $W(C_1,C_2)$.
Let us take the simplest case of non-selfintersecting loops (\fig\fauu{
A configuration of two loops}).
Then the  answer depend on whether the two loops have equal or opposite
orientations.

\leftline{{\it I. Equally oriented contours }}

$(i)$  All cylindric surfaces bounded by the two contours have
the same area,  one cut
and  two branch
points which can appear at the sites, links and cells of the overlapping
domain 2.
The configurations leading to a reducible surface (i.e., when the two
branch points can be connected by a link) are excluded from the sum.
 Let $n_0,n_1, n_2 $ be the number of sites, links and cells of
this domain.
Then the sum over the $\Omega$-factors yields
\eqn\haidebe{\eqalign{&
{n_0(n_0-1)-n_1 \over 2} \omega _{[2]}^{(0)} \omega _{[2]}^{(0)}
+ (n_0n_1-2n_1) \omega _{[2]}^{(0)}\omega _{[2]}^{(1)} \cr
&+
n_0n_2 \omega _{[2]}^{(0)}\omega _{[2]}^{(2)}+
 {1 \over 2}n_1(n_1-1)\omega _{[2]}^{(1)} \omega _{[2]}^{(1)} \cr
& +
n_1n_2 \omega _{[2]}^{(1)}\omega _{[2]}^{(2)}+
 {n_2(n_2-1) \over 2}\omega _{[2]}^{(2)}\omega _{[2]}^{(2)}\cr
&= {1\over 2}(n_0-n_1+n_2
-2\lambda n_2 )(n_0-n_1+n_2-1) +{1 \over 2}\lambda^2 n_2(n_2-1)\cr
&={1 \over 2}\lambda^2 n_2(n_2-1)\cr}
}

$(ii)$ The contact interactions  lead to a second term which is the sum over
surfaces representing two discs connected with a microscopic  tube.
Their contribution is
\eqn\ipli{n_2\omega^{(2)}_{[2]}=n_2 (-\lambda +{1 \over 2}\lambda ^2)}
The sum over the two terms gives
\eqn\woas{W(C_1,C_2)= e^{-{1\over 2}(A_1+2A_2)}(-A_2+{1\over 2}A_2^2)}

\leftline{{\it II. Oppositely  oriented contours}}

$(i)$  Besides the minimal cylindric surface with area $A_1$ there
is a whole class of cylindric surfaces with area $A_1+A_2$ (see ref.
\ivn ).
The contribution of these surfaces is (\ivn , Appendix B)
\eqn\iaiaia{e^{-{1 \over 2}A_1}+ (n_2 -1)e^{-{1 \over 2}A_1 +A_2}
 }

$(ii)$
The contact interactions
produce a term
\eqn\ttrtr{-n_2e^{-{1\over 2}A_1+A_2}}
\eqn\avac{W(C_1,C_2)= e^{-{1\over 2}(A_1-A_2)}- e^{-{1\over 2}(A_1+A_2)}}
so that the total sum depends only on the two areas
\eqn\ttotl{W(C)=-e^{-{1\over 2}A_1+A_2}+
e^{-{1 \over 2}A_1}}

\subsec{Compact lattices}
For a compact lattice $\CL$  with the topology of a sphere and area
$A_{{\rm tot}}$
the sum over surfaces becomes
much more involved.
 It will contain a
 sum over spherical surfaces
 representing  multiple branched coverings of $\CL$.
There are infinitely many
 surfaces contributing to the elementary Wilson loops. For example,
\eqn\wlplo{W_1 =  e^{-{1 \over 2}\lambda}+e^{-{1\over 2}(
A_{{\rm tot}})}+...}
Therefore, we have either to modify the weights of the irreducible
surfaces, or to use the same weights as for the infinite lattice but allow
contact interactions
due to
 microscopic
tubes, trousers, etc., connecting two, three, etc. multiplaquettes.
The amplitudes of these contact interactions are given by the
coefficients in the expansion  \asda .

We are not going to discuss this case in details since there is
nothing conceptually new.
As a single example we will calculate the two leading orders of the free
energy due to the surfaces defining
simple and double coverings of $\CL$.

Let $n_i$ be the number of $i$-cells of the lattice $\CL$ ($i=0,1,2$).
The free energy should depend only on the total area
 $A_{{\rm tot}}=n_2 \lambda $.
The two simple coverings (with the two possible orientations) contribute
the term $2e^{-{1 \over 2}A_{{\rm tot}}}$.
There are two kinds of  double coverings depending on the
orientations of the two layers.

\leftline{{\it Equal orientations}}
 \noindent
The sum over cylindric surfaces with a cut
is given by the same formula as for the two-loop correlator, with
the only difference that here the Euler relation gives $n_0-n_1+n_2
=2$. The result is $(1-\lambda n_2+{1\over 2}\lambda^2)e^{-\lambda
A_{{\rm tot}}}
$. We have to add the term with contact interaction due to the vertex
$s_{[1,1]} = (-\lambda +{1 \over 2}\lambda ^2)$
 connecting two equally oriented plaquettes:
$n_2(1-\lambda)e^{- A_{{\rm tot}}}$.

\leftline{ {\it Opposite orientations}}
\noindent
The sum over irreducible surfaces  consisting of two sheets with
opposite orientations and connected along a tree-like configuration of
links, was calculated in Appendix B of \ivn\ and
equals  $(p-2)e^{- A_{{\rm tot}}}$.
We have to add to it the term with contact interaction
due to the vertex $f_{[1,-1]} =-1$ connecting
 two plaquettes with opposite orientations: $-n_2 e^{- A_{{\rm tot}}}$.

Collecting all terms, we find
\eqn\enrg{\CF=2e^{-{1 \over 2} A_{{\rm tot}}}+
[-1-2A_{{\rm tot}} +
{1 \over 2}A^2_{{\rm tot}}]e^{- A_{{\rm tot}}}+ ...}

An interesting phenomenon on compact lattices was observed recently
by Douglas and Kazakov \kd . The sum over surfaces for the free energy on
the sphere is  convergent only  for sufficiently large total area
\eqn\hstytg{ A_{{\rm tot}} >\pi ^2}
At the critical point   $A_{{\rm tot}} =\pi ^2$
the entropy of the branched points overwhelms the  energy
 and the effective string tension vanishes.
The phenomenon resembles to what happens with the
compactified bosonic
string at the Hagedorn radius.

The singularity at $A=\pi^2$ can be observed already in
 the behaviour of the  multiple Wilson loops
surrounding a contour of area $A$.
{}From the explicit expression given in the Appendix
(with $\lambda$ replaced by
$A$) one finds the asymptotics
at  $n \to \infty$
\eqn\oppaaa{W_{n} (A)\approx {1 \over
\sqrt{\pi}}(n\sqrt{ A})^{-3/2}
\cos ( 2n \sqrt{A}-3\pi /4)}
The value of $A$ for which  the oscillations disappear is exactly the
critical  area  $A_{c}=\pi ^2$.
This is not a surprise since on a compact lattice with area $A$, the
free energy contains a term $ \sum _n W_n^2(A)$ (compare with eq. \ffru).

Finally, let make a  remark  concerning the case of
 a toroidal manifold $\CM$.
It was nicely demonstrated by Gross and Taylor that the free energy
of the gauge theory on the torus can be interpreted as the sum over all
connected nonfolding oriented
 surfaces wrapping the torus, with right combinatorial
factors. If we add to this sum the sum over microscopic surfaces \ffru ,
the result will be exactly the logarithm of the $\eta$-function
\eqn\trrs{\CF_{{\rm torus}}={1 \over 12}A_{{\rm tot}}
-\sum_{n=1}^{\infty}\log (1-e^{-n A_{{\rm tot}}})}

\appendix{B}{The one-plaquette Wilson loops}

The continuum does not depent on the choice of the one-plaquette action
which is a matter of convenience.
The simplest choice from the technical point of view is the
heat kernel action which has trivial scaling.
The Wilson loops for this action are the same as in a continuous
gauge theory defined on a disk with area $\lambda$.
The Migdal-Makeenko equations
derived in \kk yield
\eqn\looki{\eqalign{
\Big[
2{\p \over \p \lambda}&+\sum_{n\ne 0} |n| \beta _n \p_n+
 \sum_{n,k \ne 0}  \Big(
n \ \beta _n \ k \  \beta _k \ \p_{n+k} \cr
 &+{1 \over N^2}\theta (kn)
 (|n|+|k|) \beta _{n+k} \p_n\     \p_k +
\Big)\Big] e^{N^2W[\beta ]}
 =0 \cr}}
This equation is equivalent to the heat kernel equation written in the loop
space \qcdd .

In the limit $N\to \infty$ the loop equation \looki\
yields  a system of recursive equations for the connected correlators
$W_{k_1...k_n}$
\eqn\ftap{\eqalign{
(2{\p \over \p \lambda} +\sum_{i=1}^{n} k_i)W_{k_1,k_2,...,k_n}&+
\sum _{i\ne j}^{n} k_i k_{j} W_{k_i+k_j, k_2,...,\bar k_j, ...,
k_n}\cr &+\sum_{i=1}^{n}\sum _{k=1}^{k_i-1} \sum_{\alpha \in P(L)}
|k_i| W_{k_i-k, L\backslash \alpha}W_{k, \alpha}
=0\cr}}
In particular, for the simple Wilson loops
$W_{n} = (1/N)\langle \tr U \rangle$, eq. \ftap\ reads
\eqn\iioip{
2{\p \over \p \lambda}W_n+ n \sum _{k=1}^{n}W_{k}W_{n-k}=0}
This equation has been studied in
 \ros . The solution of \iioip\ is
\eqn\polop{
W_{n}=e^{-n\lambda /2}\sum_{m=0}^{n-1} {(-n\lambda)^m
\over (m+1)!}=e^{-n\lambda /2}(1-{n(n-1)\over 2}\lambda + ...)}

In the continuum limit it is sufficient to keep only the linear
term in $\lambda$.

\bigbreak\bigskip\centerline{{\bf Acknowledgements}}\nobreak
The author thanks M. Douglas, D. Gross, V. Kazakov, A. Migdal,
A. Polyakov and A. Tseytlin
for stimulating discussions, and
especially  J.-B. Zuber for his critical remarks.

\appendix{A}{ Loop equations for the puncture operators}

\subsec{  Loop equations for the link vertices}
The expansion of $F$ in terms of the momenta
\mmmi 
has been found by O'Brien and Zuber \obrz\ in the form
\odof \ 
where the coefficients $f_{[k_1, ... ,k_n]}$ were  obtained
in the large $N$ limit  by
a system of algebraic relations
suggested originally  by Kazakov in \kaza .
These relations are equivalent to  eq. \uniiy\ written in terms of loop
variables
\eqn\uniyy{
\Big( \sum_{n\ge 1}na_{n-1}\p_n+{1 \over N}
\sum_{k,n \ge 1}[ (n+k+1)a_na_k\p_{k+n+1}+ nka_{n+k-1}\p_n\p_k] -1
\Big)\ e^{F[a]}=0}
where we denoted $a_0=1, \p_n=\p / \p a_n$.
Eq. \uniyy\
is sufficient to determine   the   $1/N$ expansion of the link-vertices
$f_{[k_1,...,k_n]}$.
In the large $N$ limit the identity \uniyy\ is equivalent to a
 system of recursive relations \obrz
\eqn\oopol{\eqalign{
f_{[1]}&=1\cr
f_{[k_1,...,k_n]}+\sum _{j=2}^{n} k_j f_{[k_1+k_j,k_2,...,\bar k_j,
...,k_n]}
+\sum_{k=1}^{k_1 -1} \sum _{\alpha \in P(L)} f_{[k_1-k, L\backslash \alpha]}
f_{[k,\alpha]}&=0\cr}}
where $L={k_2,...,k_n}$ and $P(L)$ is the set of all subsets of $L$,
including the emply set, and the bar means omitting the argument below it.
The equations for the  weights $f_{[k]}$ of the disks \kaza
\eqn\plopp{f_{[1]}=1; \ \
f_{[n]}+\sum_{k=1}^{n-1} f_{[k]}f_{[n-k]}=0, \ n=1,2,...}
are solved by the Catalan numbers
\eqn\ihy{f_{[n]}=(-)^{(n-1)}{(2n-2)! \over n! (n-1)!}}

\listrefs
\listfigs
\bye